\input{aipcheck}

\documentclass[
    ,final            
  ]
  {aipproc}

\layoutstyle{8x11double}

\newcommand     \s      {\,{\rm s}}

\newcommand{\smyr}      {{ M_\odot\ \rm yr^{-1}}}
\newcommand{\sm}        {{ M_\odot}}

\newcommand{\beq}       {\begin{equation}}
\newcommand{\eeq}       {\end{equation}}
\newcommand{\beqa}      {\begin{eqnarray}}
\newcommand{\eeqa}      {\end{eqnarray}}
\newcommand{\e}         {$^{-1}$}

\newcommand{\fkep}      {f_{\rm Kep}}

\newcommand{\mds}       {\dot m_*}
\newcommand{\msd}       {m_{*d}}
\newcommand{\msdo}       {m_{*d,\, 0}}
\newcommand{\mdsd}      {\dot m_{*d}}
\newcommand{\mdsdo}      {\dot m_{*d,\, 0}}
\newcommand{\tff}       {t_{\rm ff}}

\newcommand{\htwo}      {H$_2$}

\newcommand{\lal}       {Lyman-$\alpha$}

\newcommand{\prad}      {{P}_{\rm rad}}

\newcommand{\pbyp}[2]   {\frac{\partial #1}{\partial #2}}

\newcommand{\esd}       {\epsilon_{*d}}
\newcommand{\esdb}       {\bar\epsilon_{*d}}
\newcommand{\fsh}       {f_{\rm sh}}

\newcommand{\mst}       {m_{*,\,2}}
\newcommand{\msdt}      {m_{*d,\,2}}

\newcommand{\kapr}      {\kappa_{\rm R}}

\newcommand{\ledd}      {L_{\rm Edd}}

\newcommand{\teff}      {T_{\rm eff}}

\newcommand{\phiedd} {\phi_{\rm Edd}}
\def\ion#1#2{#1$\;${\small\rm II}\relax}
\def\lesssim{\mathrel{\hbox{\rlap{\hbox{\lower4pt\hbox{$\sim$}}}\hbox{$<$}}}}
\def\gtrsim{\mathrel{\hbox{\rlap{\hbox{\lower4pt\hbox{$\sim$}}}\hbox{$>$}}}}

\newcommand{\apj}       {\emph{Apj}}

\begin{document}

\title{Star Formation at Zero and Very Low Metallicities}


\author{Jonathan C. Tan}{
  address={Department of Astronomy, University of Florida, Gainesville, FL 32611, USA.\\jt@astro.ufl.edu}
}

\author{Christopher F. McKee}{
  address={Departments of Physics \& Astronomy, University of California,
Berkeley, CA 94720, USA.\\cmckee@astro.berkeley.edu}
}

\begin{abstract}
We describe how star formation is expected to proceed in the early
metal-free Universe, focusing on the very first generations of
stars. We then discuss how the star formation process may change as
the effects of metallicity, external radiative feedback, and magnetic
and turbulent support of the gas become more important. The very first
stars (Pop III.1) have relatively simple initial conditions set by
cosmology and the cooling properties of primordial gas. We describe
the evolution of these stars as they grow in mass by accretion from
their surrounding gas cores and how the accretion process is affected
and eventually terminated by radiative feedback processes, especially
\ion{H}{2} region expansion and disk photoevaporation. The ability of
the protostar and its disk to generate dynamically important magnetic
fields is reviewed and their effects discussed.  Pop III.1 star
formation is likely to produce massive ($\sim 100-200\sm$) stars that
then influence their surroundings via ionization, stellar winds, and
supernovae. These processes heat, ionize and metal-enrich the gas,
thus altering the initial conditions for the next generation of star
formation.  Stars formed from gas that has been altered significantly
by radiative and/or mechanical feedback, but not by metal enrichment
(Pop III.2) are expected to have significantly smaller masses than Pop
III.1 stars because of more efficient cooling from enhanced HD
production. Stars formed from gas that is metal-enriched to levels
that affect the dynamics of the collapse (the first Pop II stars) are
also expected to have relatively low masses. We briefly compare the
above star formation scenarios to what is known about present-day star
formation.
\end{abstract}

\maketitle


\section{Introduction}

The first generation of stars is expected to have played a
significant role in reionizing the Universe, for which observational
constraints can be derived from CMB polarization \cite{page2007} and
future high redshift 21~cm HI observations \cite{morales2004}. The
first stars polluted their pristine surroundings with metals and may
have contributed to the observed metal abundances of Galactic halo
stars
\cite{beers2005} and the Ly-$\alpha$ forest \cite{schaye2003,norman2004}.
Light from the first stars may contribute to the observed NIR
background intensity, e.g. \cite{santos2002,fernandez2006}, and its
fluctuations \cite{kashlinsky2004} (however, see \cite{thompson2007}). 
If massive, the deaths of the first stars may be observable as
supernovae \cite{weinmann2005} or gamma-ray bursts \cite{bromm2002}.
The first stars thus set the stage for galaxy formation, for which
observational constraints now exist at redshifts 8 to 10
\cite{stark2007}, and supermassive black hole formation and growth,
with $\sim$billion solar mass black holes being seen out to redshift
6.4 \cite{fan2003,willott2003}.

In discussing the formation of the first stars, the terms ``first
stars'' and ``Population III stars'' are often used interchangeably,
but this can lead to confusion. To be precise, following McKee \& Tan
\cite{mckee2008}, we shall define Population III stars as those stars
with a metallicity sufficiently low that it has no effect on either
the formation or the evolution of the stars. The value of the critical
metallicity for star formation---i.e., the value below which the
metals do not influence star formation, e.g. the initial mass
function (IMF)---is uncertain, with estimates ranging from $\sim
10^{-6}Z_\odot$ if the cooling is dominated by small dust grains that
contain a significant fraction of the metals
\cite{omukai2005} to $\sim 10^{-3.5}Z_\odot$ if the cooling is
dominated by the fine structure lines of C and O \cite{bromm2003}. It
is possible that even lower values of the metallicity could
significantly affect the evolution of primordial stars (Meynet,
private communication).  Among Population III stars, we distinguish
between the first and second generations, which we term Population
III.1 and III.2, respectively.  We define Population III.1 stars as
those for which the initial conditions are determined solely by
cosmological fluctuations. By definition, the collapse of gas to form
these stars is mediated by the cooling properties of primordial or
near primordial composition gas that has not been significantly
influenced by stellar or AGN radiation fields. Since we expect the
effects of radiation to be more pervasive than the spread and mixing
of metals, the composition of Pop III.1 halos should in fact be
precisely that resulting from big bang nucleosynthesis.

As discussed later in this article, the initial conditions for
Population III.2 stars are significantly affected by other stars,
primarily by radiation.  Lyman-Werner band radiation destroys \htwo\
molecules, reducing the cooling efficiency of the gas.  Ionizing
radiation heats up the gas to temperatures $\sim 25,000$~K
\cite{giroux1996,shapiro2004}, depending on the temperature of the
radiation field. While ionized, thermal expansion can drive the gas
out of low-mass dark matter halos. Shocks associated with \ion{H}{2}
regions and supernova remnants can also ionize and displace the gas.
If the gas has a chance to recombine, the relatively high residual
electron fraction catalyzes molecule formation, particularly HD
formation, which can dramatically enhance subsequent cooling, perhaps
reducing the characteristic star-formation mass\cite{uehara2000}.
Greif \& Bromm\cite{greif2006} use the term ``Population II.5" to
describe stars that form from gas in which HD cooling is important,
whereas in our terminology these would be Population III.2 stars. We
believe that it is better to describe all essentially metal-free stars
as ``Population III." It should be noted that this definition of
Population III.2 stars includes all Population III stars that were
significantly affected by previous generations of star formation, even
if that did not result in significant HD production.  According to
Greif \& Bromm \cite{greif2006}, Population III.1 stars are relatively
rare, constituting only about 10\% by mass of all Population III
stars.  Note, however, that the precise intensity of the background
radiation field that causes a transition from Pop III.1 to Pop III.2
remains to be determined from numerical calculations.

\section{Population III.1 Star Formation}

\subsection{Initial Conditions and Accretion Rate}

The initial conditions for the formation of the first stars are
thought to be relatively well understood: they are determined by the
growth of small-scale gravitational instabilities from cosmological
fluctuations in a cold dark matter universe. The first stars are
expected to form at redshifts $z\sim 10-50$ in dark matter
``minihalos'' of mass $\sim 10^6 M_\odot$ \cite{tegmark1997}.  In the
absence of any elements heavier than helium (other than trace amounts
of lithium) the chemistry and thermodynamics of the gas are very
simple. Once gas collects in the relatively shallow potential wells of
the minihalos, cooling is quite weak and is dominated by the
ro-vibrational transitions of trace amounts of $\rm H_2$ molecules
that cool the gas to $\sim 200$~K at densities $n_{\rm H} \sim
10^4\:{\rm cm^{-3}}$ \cite{abn2002,bcl2002}. As the gas core contracts
to greater densities, the cooling becomes relatively inefficient and
the temperature rises to $\sim 1000$~K. At densities $\sim
10^{10}\:{\rm cm^{-3}}$ rapid 3-body formation of $\rm H_2$ occurs,
creating a fully molecular region that can cool much more
efficiently. This region starts to collapse supersonically until
conditions become optically thick to the line and continuum cooling
radiation, which occurs at densities $\sim 10^{17}\:{\rm cm^{-3}}$.
Recent 3D numerical simulations have advanced to densities of order
$10^{21}$~cm$^{-3}$ (see contributions by Turk et al. and Yoshida et
al., these proceedings), but have trouble advancing further given the
short simulations timesteps required to resolve the dynamics of the
high density gas of the protostar. Further numerical progress can be
achieved by introducing sink particles \cite{bromm2004}.

We have adopted the alternative approach of modeling the protostar's
accretion and evolution analytically \cite{tan2004}. The accretion
rate depends on the density structure and infall velocity of the gas
core at the point when the star starts to form. Omukai \& Nishi
\cite{omukai1998} and Ripamonti et al. \cite{ripamonti2002} showed
that the accreting gas is isentropic with an adiabatic index
$\gamma\simeq 1.1$ due to H$_2$ cooling; i.e., each mass element
satisfies the relation $P=K\rho^\gamma$ with the ``entropy parameter''
$K=$~const.  In hydrostatic equilibrium---and therefore in a subsonic
contraction---such a gas has a density profile $\rho \propto
r^{-k_\rho}$ with $k_\rho\simeq 2.2$.
This is confirmed by numerical simulations, which show
that just before protostar formation the gas at the center of the
minihalo has organized itself into an approximately singular
polytropic sphere (with a modest degree of flattening about a rotation
axis) with $k_\rho=2.2$. 
In our model, we 
describe the normalization of the density
structure of core via the entropy parameter
\beq
K'\equiv \frac{P/\rho^\gamma}{1.88\times 10^{12}~{\rm cgs}}=
         \frac{\teff'}{300~{\rm K}}\left(\frac{10^4{\rm cm}^{-3}}{n_{\rm H}}\right)^{0.1},
\label{eq:kp}
\eeq
where $\teff' \equiv T+\mu\sigma^2_{\rm turb}/k$ is an effective
temperature that includes the modest effect of subsonic turbulent motions
that are seen in numerical simulations \cite{abn2002}.

Simulations show the gas is inflowing subsonically at about a third of
the sound speed \cite{abn2002}. We thus use Hunter's \cite{hunter1977}
solution for mildly subsonic inflow (Mach number =0.295), with a
density that is 1.189 times greater than a singular isothermal sphere
at $t=0$, and the accretion rate is 2.6 times greater than the Shu
\cite{shu1977} solution.  

Feedback from the star, whether due to winds, photoionization, or
radiation pressure, can reduce the accretion rate of the star. We
define a hypothetical star$+$disk mass, $\msdo$, and accretion rate,
$\mdsdo$, in the absence of feedback. In this case, the star$+$disk mass
equals the mass of the core out of which it was formed, $\msdo=M(r)$.
The instantaneous and mean star formation efficiencies are
\beq
\esd\equiv\frac{\mdsd}{\mdsdo}, 
\label{eq:esd}
\eeq
\beq
\esdb \equiv \frac{\msd}{\msdo}=\frac{\msd}{M}.
\label{eq:esdb}
\eeq

Assuming the Hunter solution applies for a singular polytropic sphere
with $\gamma=1.1$, the accretion rate is then
\beq
\label{eq:mdot}
\mdsd=0.026 \epsilon_{*d}K'^{15/7} (M/\sm)^{-3/7}~M_\odot~{\rm yr}^{-1},
\eeq
with the stellar mass smaller than the initial enclosed core mass via
$m_* \equiv \msd/(1+f_d) =\esdb M/(1+f_d)$. We choose a fiducial value
of $f_d=1/3$ appropriate for disk masses limited by enhanced viscosity
due to self-gravity.  We compare this analytic accretion rate (for the
case of no feedback) with numerical estimates in Fig. 1.

\begin{figure}
  \includegraphics[height=.3\textheight]{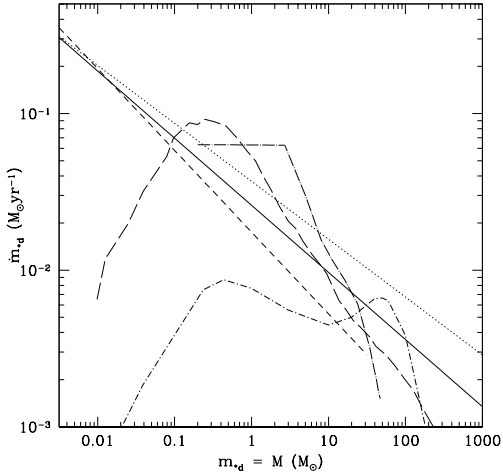} \caption{Mass
  accretion rate onto the protostar$+$disk as a function of their
  total mass $m_{*d}$ for the case of negligible stellar
  feedback. {\it Solid} line is the fiducial model from Tan \& McKee
  \cite{tan2004} (with $K^\prime=1$) from eq. (\ref{eq:mdot}).  {\it
  Dotted} line is from the 1D model of Omukai \& Nishi
  \cite{omukai1998}. {\it Dashed} line is the analytic result from
  Ripamonti et al. \cite{ripamonti2002}. {\it Dot-dashed} line is the
  settling inflow rate at the final stage of the simulation of Abel et
  al. \cite{abn2002}, now as a function of the enclosed mass. Note
  that the decline of this rate at small masses is due to the lack of
  the full set of high density cooling processes in their
  simulation. {\it Long-dashed} line is the equivalent quantity from
  Yoshida et al.\cite{yoshida2006}. {\it Dot-long-dashed} line is the
  sink particle accretion rate of Bromm \& Loeb \cite{bromm2004}.  }
\end{figure}

O'Shea \& Norman \cite{o'shea2007} studied the properties of Pop III.1
pre-stellar cores as a function of redshift. They found that cores at
higher redshift are hotter in their outer regions, have higher free
electron fractions and so form larger amounts of $\rm H_2$ (via $\rm
H^-$), although these are always small fractions of the total mass. As
the centers of the cores contract above the critical density of
$10^4\:{\rm cm^{-3}}$, those with higher $\rm H_2$ fractions are able
to cool more effectively and thus maintain lower temperatures to the
point of protostar formation. The protostar thus accretes from
lower-temperature gas and the accretion rates, proportional to $c_s^3
\propto T^{3/2}$, are smaller. Measuring infall rates at the time of
protostar formation at the scale of $M=100\sm$, O'Shea \& Norman find
accretion rates of $\sim 10^{-4}\smyr$ at $z=30$, rising to $\sim
2\times 10^{-2}\smyr$ at $z=20$. This corresponds to a range in
$K^\prime$ of 0.20 to 2.4. However, for Hunter's mildly subsonic
solution, which we have suggested is closest to the numerical
simulations, the accretion rate increases from $0.70 c_s^3/G$ at large
radii ($r\gg c_s t$, where $t=0$ is the moment of protostar formation)
to $2.58c_s^3/G$ at small radii (Hunter 1977), an increase of a factor
3.7. This demonstrates that caution should be exercised in inferring
accretion rates at late times from those measured at early times.

Spolyar et al. \cite{spolyar2007} considered the effect of heating due
to WIMP dark matter annhilation on the collapse of Pop III.1
pre-stellar cores. In their fiducial model of the dark matter density
profile and for a WIMP mass of 100~GeV, heating dominates cooling (as
evaluated by Yoshida et al. \cite{yoshida2006}) for $n_{\rm
H}>10^{13}\:{\rm cm^{-3}}$, i.e. in the inner $\sim 20$~AU. In this
case the collapse can be expected to be halted inside this region and
a quasi-hydrostatic object created, although detailed models of the
dynamics remain to be worked out. However, as more baryons from the
outer regions cool and join this central core, the luminosity needed
to support it increases rapidly with mass; 
the increase provided by WIMP annihilation is uncertain.  However, in
view of the fact that most of the later accretion of baryons should
occur through a disk (see below), which would tend to enhance the
baryon to dark matter mass ratio in the central disk and star, we
expect any heating due to WIMP annhilation will diminish in importance
as the stellar mass grows.
It should be noted that the WIMP annhilation heating rate depends
sensitively on the density profile of the dark matter, which is so far
not adequately resolved in numerical simulations of Pop III.1 core
formation.

\subsection{Protostar and Accretion Disk Evolution}

Assuming dark matter annhilation does not have a significant effect on
the main phase of protostellar accretion and noting that there may be
a range of accretion rates depending on the formation redshift
\cite{o'shea2007}, we now consider what happens to the collapsing gas
as it forms a protostar and accretion disk at rates that are initially
$\gtrsim 10^{-2}\smyr$ (i.e. eq.~\ref{eq:mdot} with $K^\prime\sim1$).

Rotation of the infalling gas has a dramatic effect on the evolution
of the protostar, since it leads to lower gas densities and optical
depths in the regions near the stellar surface that are above and
below the disk. This leads smaller photospheric radii and thus higher
temperature radiation fields that can then have a stronger dynamical
influence on the infall. Following the treatment of Tan \& McKee
\cite{tan2004}, we parameterize the rotation in terms of
\beq
\fkep\equiv\frac{v_{\rm rot}(r_{\rm sp})}{v_{\rm Kep}(r_{\rm sp})},
\label{eq:fkep}
\eeq
the ratio of the rotational velocity to the Keplerian velocity
measured at the sonic point at $r_{\rm sp}$. Averaging in spherical
shells, Abel, Bryan, \& Norman
\cite{abn2002} found $\fkep\sim 0.5$ independent of radius, so we
adopt this as a fiducial value. If angular momentum is conserved
inside the sonic point, then the accreting gas forms a disk with an
outer radius
\beqa
r_d &=& 1280\left(\frac{\fkep}{0.5}\right)^2\left(
        \frac{\msdt}{\esdb}\right)^{9/7}K'^{-10/7}~{\rm AU},\nonumber\\
        &\rightarrow& 1850\left(\frac{\fkep}{0.5}\right)^2
        \frac{\mst^{9/7}}{K'^{10/7}}~{\rm AU},
\label{eq:rd}
\eeqa
where the $\rightarrow$ is for the case with $f_d=1/3$.

The first gas to collapse has a small disk-circularization radius and
falls directly on to the protostar.  The size of the protostar depends
on the accretion rate during its formation history.  At lower masses
there is a balance in the size set by the need to radiate the
luminosity, which is mostly due to accretion, with a photospheric
temperature that is likely to be close to $\sim 6000$~K because the
opacity due to $\rm H^-$ rapidly declines below this temperature.
Under the assumption of spherical accretion, Stahler, Palla, \&
Salpeter \cite{stahler1986} found the protostellar radius to be
\beq
\label{stahler}
r_* \simeq 67 (m_*/\sm)^{0.27} \dot{m}_{*,-2}^{0.41}~~~ R_\odot,
\eeq 
where $\dot{m}_{*,-2}\equiv \dot{m}_*/(10^{-2}\;\smyr)$.  For the
accretion rates typical of primordial star formation we see that the
size is large, but small compared to the disk radius for masses
$\gtrsim \sm$.

\begin{figure}
  \includegraphics[height=.3\textheight]{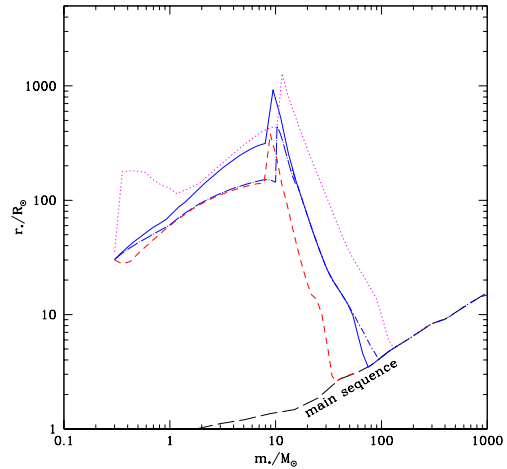}
  \caption{\label{fig:r} Evolution of protostellar radius with mass,
  based on the model of Tan \& McKee \cite{tan2004}. All models shown
  have a rotation parameter $f_{\rm Kep}=0.5$. The {\it solid} line
  shows the evolution of the fiducial model with $K^\prime=1$, a
  Shakura-Sunyaev disk viscosity parameter of $\alpha_{\rm ss}=0.01$,
  and a reduction of accretion rate due to feedback effects
  \cite{mckee2008}, which becomes important for $m_*\gtrsim50\sm$. The
  star joins the main sequence \cite{schaerer2002}, shown by the {\it
  long dashed} line, at about $80\sm$. The {\it dot-dashed} line,
  visible only from $50-100\sm$, shows the same model but with no
  reduction in accretion rate due to feedback effects. The {\it
  dot-long-dashed} line, visible up to about $20\sm$ shows the
  fiducial model, but with $\alpha_{\rm ss}=0.3$ (appropriate for
  viscosity driven by self-gravity \cite{gammie2001}). The {\it
  dotted} line shows the fiducial model but with $K^\prime=2$, while
  the {\it dashed} line shows the $K^\prime=0.5$ case.
}
\end{figure}

As the star grows in mass its internal luminosity increases rapidly
and begins to dominate over accretion luminosity
\cite{omukai2003,tan2004}. Once the star is older than its
instantaneous Kelvin-Helmholz time (i.e. its gravitational energy
divided by its internal luminosity), it begins to contract towards the
main sequence configuration. At this time there is an associated
temporary swelling of the outer layers of star by a factor of about
two due to structural rearrangements inside the star.  According to
Schaerer \cite{schaerer2002}, the main sequence radius for nonrotating
stars is
\beq
r_*\simeq 4.3\mst^{0.55}~R_\odot~~~~~({\rm main\ sequence})
\eeq
to within 6\% for $0.4\leq\mst\leq 3$.  If the star continues to
accrete and gain mass, it will approximately follow this mass-size
relation, although rotation will tend to swell the equatorial surface
layers somewhat. The evolution of the protostellar radius is shown for
several models in Fig.~\ref{fig:r}

The high accretion rates of primordial protostars make it likely that
the disk will build itself up to a mass that is significant compared
to the stellar mass. At this point the disk becomes susceptible to
global ($m=1$ mode) gravitational instabilities
\cite{adams1989,shu1990}, which are expected to be efficient at
driving inflow to the star, thus regulating the disk mass. Thus we
assume a fixed ratio of disk to stellar mass, $f_d=1/3$, in our
models.

Accretion through the disk may also be driven by local instabilities,
the effects of which can be approximated by simple Shakura-Sunyaev
$\alpha_{\rm ss}$-disk models. Two dimensional simulations of clumpy,
self-gravitating disks show self-regulation with $\alpha_{\rm
ss}\simeq (\Omega t_{\rm th})^{-1}$ up to a maximum value $\alpha_{\rm
ss}\simeq 0.3$ \cite{gammie2001}, where $\Omega$ is the orbital
angular velocity, $t_{\rm th}\equiv \Sigma kT_{\rm c,d}/(\sigma T_{\rm
eff,d}^4)$ is the thermal timescale, $\Sigma$ is the surface density,
$T_{\rm c,d}$ is the disk's central (midplane) temperature, and
$T_{\rm eff,d}$ the effective photospheric temperature at the disk's
surface. Fragmentation occurs when $\Omega t_{\rm th}\lesssim 3$: this
condition has the best chance of being satisfied in the outermost
parts of the disk that are still optically thick. However, Tan \&
Blackman
\cite{tan2004b} considered the gravitational stability of constant 
$\alpha_{\rm ss}=0.3$ disks fed at accretion rates given by
eq. \ref{eq:mdot} and found that the optically thick parts of the disk
remained Toomre stable ($Q>1$) during all stages of the growth of the
protostar. We therefore expect that most Pop III.1 accretion disks, at
least in their early stages, will grow in mass and mass surface
density to the point at which gravitational instabilities, both global
and local, then mediate accretion to the star.

In addition to gravitational instabilities, the magneto-rotational
instability (MRI) may develop if dynamically important
magnetic fields are generated by dynamo action in the disk (see
below). This process could be particularly important in the inner
accretion disk, where gravitational instability is suppressed. The MRI
is thought to yield viscous stresses that correspond to somewhat
smaller values of $\alpha_{\rm ss}\sim 0.01$
\cite{balbus1998}, although the actual effective value remains unclear.

\subsubsection{Generation of Magnetic Fields in the Accretion Disk}

Given the right conditions, dynamo action can amplify a seed magnetic
field. The basic requirement is that the field is frozen into the gas
and that gas motions repeatedly stretch the field, e.g. via turbulent
or shearing motions. Here we briefly describe the results of Tan \&
Blackman \cite{tan2004b}, who considered these issues.

The seed field for the primordial star-forming environment can be
estimated from the Biermann battery mechanism for non-barotropic
flows.  On scales much larger than that of an individual star,
numerical simulations indicate that by $z=18$, seed fields of order
$\bar{B}_0\sim 10^{-26}$ Gauss can be generated \cite{kulsrud1997}.
We assume such a field is generated inside the region around the
minihalo on scales $\sim 100\:{\rm pc}$. The infall speeds within this
region are of order 1 km~s$^{-1}$, vastly greater than the drift
speeds of the ions through the neutrals (the ionization fractions are
much greater than the equilibrium values for the gas, which has a
minimum temperature of $\sim 200\:{\rm K}$). Thus the field is frozen
into the contracting gas: going from parsec to AU scales increases the
field strength by a factor of $\sim 10^{10}$, so the seed field in the
disk may then be $\sim 10^{-16}\:{\rm G}$. This estimate is highly
uncertain. Seed fields several orders of magnitude larger could arise
in the protostellar disk from the Biermann battery term if the
pressure and density gradients are misaligned inside this region as
well. A further complication that could lead to lower field strengths
is the possibility of efficient turbulent diffusion of the field so
that flux freezing and field advection do not play as significant a
role. However, since the disk-dynamo amplifies the initial field
exponentially, the uncertainties in the strength of the seed field do
not significantly affect the time to reach saturation.

A seed field is expected to be amplified by a combination of shearing
and turbulent motions in the disk. As discussed above, initially the
turbulence is thought to be due to self-gravity \cite{gammie2001}, but
as the field strength grows, this could be replaced by the MRI. The
field strength grows exponentially with a growth time $\sim
\alpha_{\rm ss}^{-1/2}\Omega^{-1}$. Field growth stops when the
fields become dynamically important, i.e. for $B\sim
(4\pi\rho)^{1/2}\alpha_{\rm ss}c_s$. The generation of strong B-fields
that are ordered on scales large compared to the disk thickness
requires turbulence in a stratified disk: the helical disk
dynamo\cite{blackman2002}. More specifically the turbulence must
generate net helicity in each hemisphere. If this occurs, and
numerical simulations are required for confirmation, then these
large-scale B-fields could start to drive a hydromagnetic bipolar disk
wind, whose feedback effects will be discussed below.

The presence of strong B-fields in the disk and their possible
advection into the protostar, then allows a star-disk interaction that
can help the star to shed angular momentum and may set the rotation
rate with which it is born. Note that models of primordial
protostellar evolution predict that the star is mostly radiatively
stable in the pre-H nuclear burning phase, which implies there is very
little dynamo amplification of B-field in the star independent of the
disk.

\subsubsection{Radial Structure of the Accretion Disk}

Given these considerations, a first approximation for the radial
structure of the protostellar accretion disk can be made by using the
standard theory of steady, thin, viscous accretion disks, with a
spatially constant viscosity parameter, $\alpha_{\rm ss}$, and for
simplicity ignoring energy injection from the star. Tan \& McKee
\cite{tan2004} calculated the evolution of the radial structure, 
assuming the disk is fed at a rate given by equation (\ref{eq:mdot})
and the inner boundary is at the protostellar surface, $r_*$. The viscosity
is assumed to be a function of the total pressure. Typically gas
pressure dominates over radiation pressure in the earlier stages
($m_*\lesssim 20\sm$). We chose $\alpha_{\rm ss}=0.01$ as a fiducial
value, that may arise from viscosity provided by the magneto
rotational instability \cite{balbus1998}.  

The radial structure of the disk is governed by the equations of
energy conservation and of angular momentum conservation.  Energy
conservation gives the emergent flux as
\beqa
F_0 &=& \frac{\mds}{4\pi\varpi}\frac{\partial}{\partial\varpi}
        \left(\frac 53 \bar\epsilon_{\rm th}+\bar\epsilon_I\right)
        +\frac{3Gm_*\mds f}{8\pi\varpi^3},\\
    & \equiv & \phi_I\left( \frac{3Gm_*\mds f}{8\pi\varpi^3}\right).
\label{eq:en}
\eeqa
Here 
\beq
f\equiv 1-\left(\frac{\varpi_0}{\varpi}\right)^{1/2},
\label{eq:f}
\eeq
is the factor that embodies the boundary condition that angular
momentum cannot be transferred across a surface on which the angular
velocity has no gradient; $\varpi_0$ is the cylindrical radius at
which $\partial\Omega/\partial\varpi$ vanishes, which we take to be
equal to the stellar radius.  The dimensionless factor $\phi_I$
describes the advection of thermal and internal energy in the disk and
is generally less than unity.

       To evaluate the angular momentum transfer in the disk, we adopt
the $\alpha$-disk model of Shakura \& Sunyaev \cite{shakura1973}, in
which the transverse stress in the disk is proportional to the
pressure, $w_{\varpi\phi}=-\frac 32\alpha P$ (we have included the
factor $\frac 32$ to conform with the convention of Frank et
al. \cite{frank1995}).  The equation describing angular momentum
transport is then
\beq
\mds\Omega f=6\pi\alpha\int_0^{z_s} Pdz,
\label{eq:angmom}
\eeq
where $z_s$ is the height of the surface of the disk.

\begin{figure}
  \includegraphics[height=0.7\textheight]{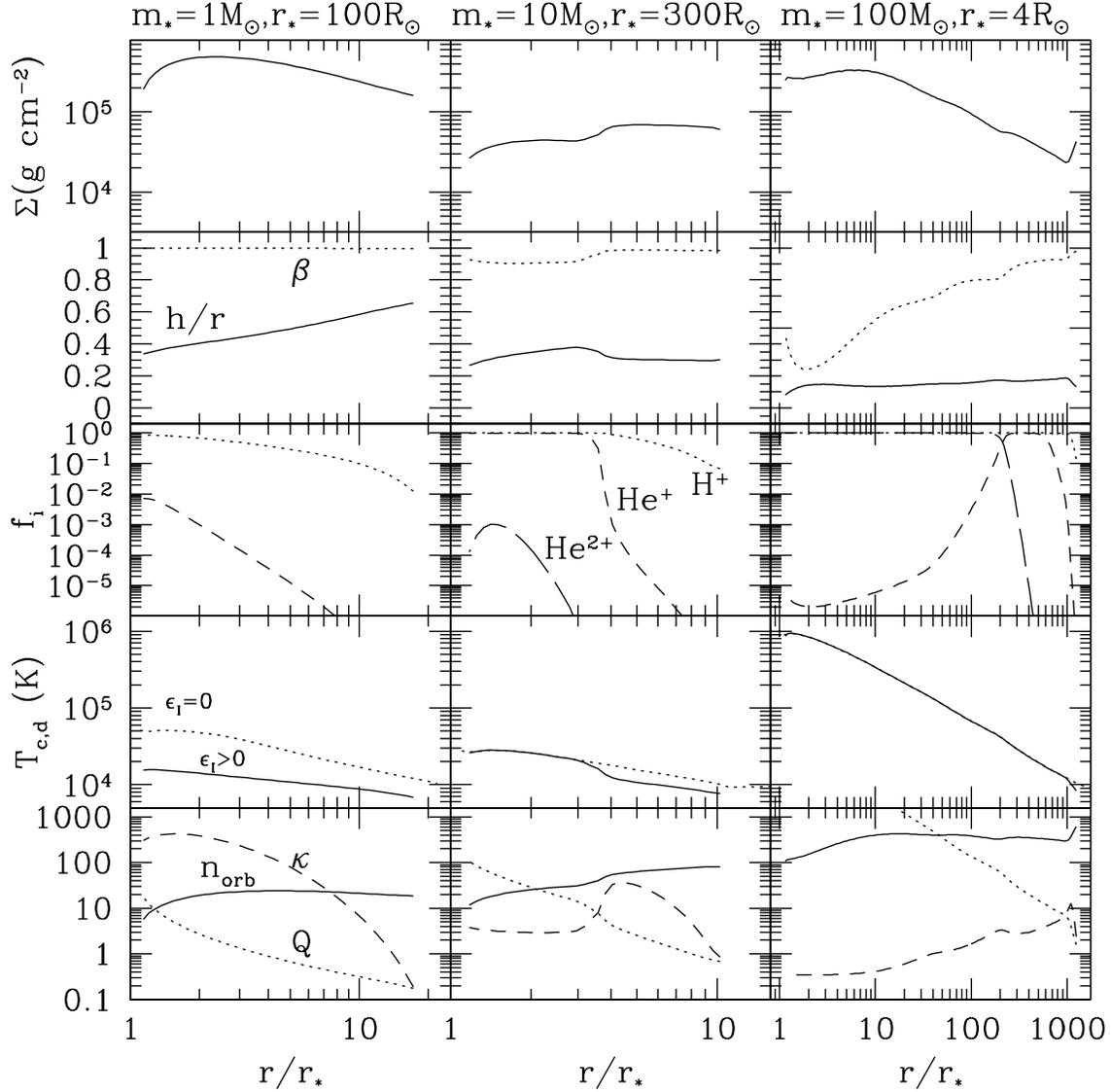}
\caption{
\label{fig:disk}
Protostellar disk structure \cite{tan2004b} for models with
$\alpha_{\rm ss}=0.01$ and $m_*=1,10,100\sm$, for which
$r_*=100,300,4R_\odot$ and $\mds=(17,6.4,2.4)\times 10^{-3}\smyr$,
respectively, i.e. from eq. \ref{eq:mdot} with no feedback. From top
to bottom the panels show (1) surface density, $\Sigma$; (2) ratio of
scaleheight to radius, $h/r$, and ratio of gas pressure to total
pressure, $\beta$; (3) midplane ionization fractions of $\rm
H^+,\:He^+,\:He^{2+}$; (4) disk midplane temperature, $T_{\rm c,d}$
(the dotted lines show results for when the ionization energy is
neglected); (5) number of orbits before accretion to the star, $n_{\rm
orb}$, Toomre $Q$ stability parameter, and Rosseland mean opacity
$\kappa$, evaluated at the midplane.  Note that all quantities are
azimuthal and temporal averages of the disk, which, being turbulent,
would exhibit local fluctuations.  }
\end{figure}

The radial disk structure at three different stages
($m_*=1,10,100\sm$) is shown in Fig.~\ref{fig:disk}, with the radial
coordinate measured in units of the protostellar radius.

\subsubsection{Vertical Structure of the Accretion Disk}

McKee \& Tan \cite{mckee2008} calculated the vertical structure of
Pop III.1 accretion disks. The vertical structure of the disk is
governed by three equations: first is the first moment of the
radiative transfer equation,
\beq
\pbyp{\prad}{z}=-\frac{\rho\kappa_F F}{c},
\label{eq:firstmom}
\eeq
where $\kappa_F$ is the flux-weighted mean opacity per unit mass and
$F(z)$ is the radiative flux. We assume that the effective optical
depth for true absorption, $\tau^*=(\tau_{\rm abs}\tau_{\rm
scatt})^{1/2}$, is much greater than unity so that the gas is
approximately in LTE \cite{shakura1973, artemova1996}. Then
$\prad\simeq\frac 13 aT^4$ and $\kappa_F\simeq \kapr$, where $\kapr$
is the Rosseland mean opacity per unit mass, so that equation
(\ref{eq:firstmom}) reduces to the equation of radiative diffusion,
\beq
\pbyp{T}{z} = -\frac{3 \kapr \rho F}{16 \sigma T^3}.
\label{eq:dtdz}
\eeq

    The second equation describes the growth of the flux due to
viscous dissipation \cite{shu1992},
\beq
\pbyp{F}{z}=-\phi_I w_{\varpi\phi}\varpi\;\pbyp{\Omega}{\varpi}
        =\frac 94 \phi_I\alpha\Omega P.
\label{eq:fz}
\eeq
We have included the factor $\phi_I$ to allow for the reduction in the
flux by the advection of internal energy.  In addition to the factor
$\frac 32\phi_I$, equation (\ref{eq:fz}) differs from the expression
adopted by Shakura \& Sunyaev \cite{shakura1973} in that it has
$\partial F/\partial z\propto P$ rather than $\propto\rho$. One can
show, however, that the height of the disk is very insensitive to this
change.  Integration of equation (\ref{eq:fz}) together with equation
(\ref{eq:angmom}) leads directly to the energy equation (\ref{eq:en}).

Finally, we have the equation of hydrostatic equilibrium,
\beq
\pbyp{P}{z}
        =-\frac{\rho Gm_*z}{\varpi^3},
\eeq
where the pressure $P$ includes both gas pressure and radiation pressure,
\beq
P=P_g+\prad=\frac{\rho kT}{\mu}+\frac{4\sigma T^4}{3c}.
\eeq

\begin{figure}
  \includegraphics[height=0.3\textheight]{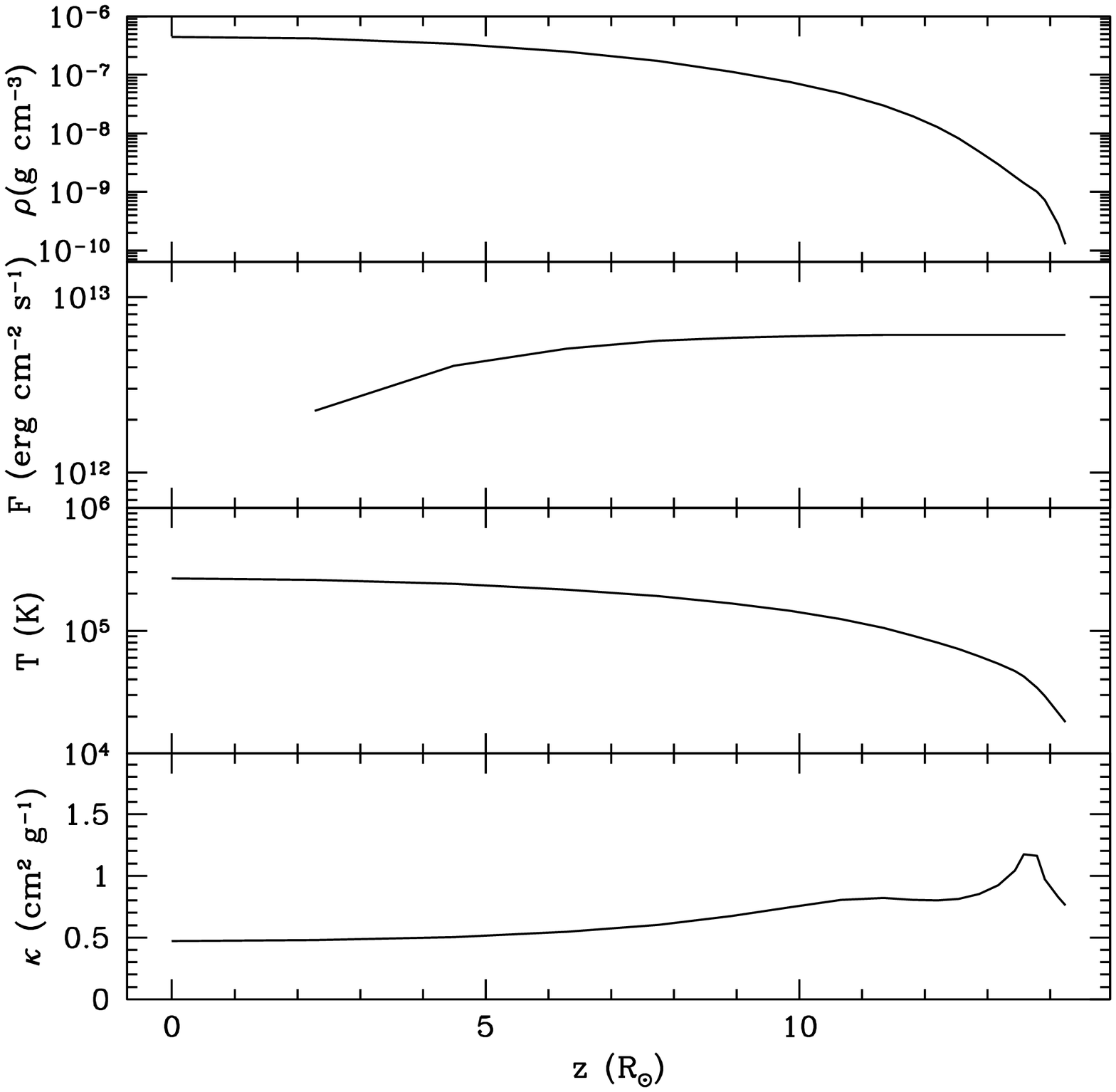}
\caption{
\label{fig:struct}
Vertical structure of the accretion disk at $r=10r_*\simeq 43R_\odot$ for
$m_*=100\:M_\odot$, $K^\prime=1$, $f_{\rm Kep}=0.5$ and no reduction
in accretion efficiency due to feedback.  }
\end{figure}

A numerical solution of these equations using the opacities of
Iglesias \& Rogers \cite{iglesias1996} is shown in
Fig.~\ref{fig:struct} for one particular stage of the protostellar
evolution for a location at 10$R_*$ around a 100$\sm$ main sequence
star, accreting at $2.4\times 10^{-3}\:M_\odot\:{\rm yr^{-1}}$,
i.e. the fiducial rate from a $K'=1$ core with no reduction due to
feedback.

More generally, we follow the disk structure during the course of the
protostellar evolution (i.e. as $m_*$, $\dot{m}_*$ and $r_*$ evolve).
As we discuss below, the disk thickness is one important factor in
determining the stellar mass at which accretion is shut off. At masses
$m_*\sim 100\sm$ we find the aspect ratio of the disk surface,
$z_s/r$, is about twice $h/r$, where $h$ is the scale height,
and has a very weak dependence on radius.

\subsection{Feedback Processes and the IMF}

\subsubsection{Radiative Feedback}

\begin{figure}
  \includegraphics[height=0.3\textheight]{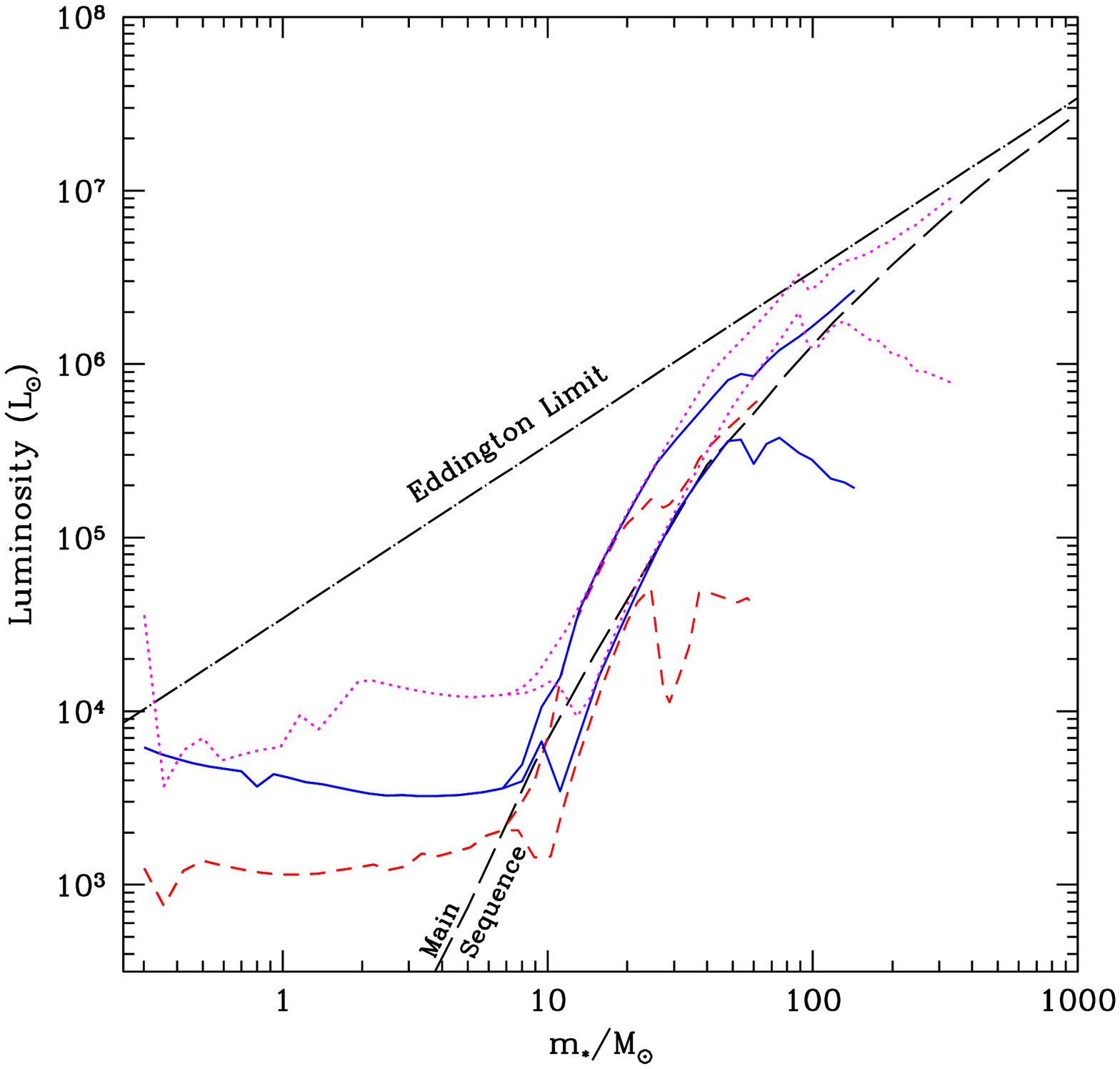}
\caption{
  \label{fig:Lstarreview} Protostellar bolometric luminosities for
models with $K^\prime=0.5,1,2$ ({\it dashed, solid, dotted} lines). In each
case the total luminosity is shown with the higher line and the
accretion luminosity (boundary layer $+$ inner disk) is shown with the
lower line. The Eddington limit is indicated with the {\it
dot-long-dashed} line and the zero age main sequence luminosity
\cite{schaerer2002} with the {\it long-dashed} line.  }
\end{figure}

The evolution of the bolometric luminosity of the protostar and the
inner accretion disk is shown in Fig.~\ref{fig:Lstarreview} for three
different accretion rates corresponding to $K^\prime=0.5,1,2$. The
luminosities are sub-Eddington for almost the entire evolution. Only
the $K^\prime=2$ model (with the highest accretion rate) exceeds the
Eddington limit: at the low-mass limit this is due to the initial
condition having a stellar radius that is too small and the star
quickly adjusts by undergoing rapid expansion; at about $80\sm$ the
star is close to joining the main sequence and so has a relatively
small size and is still accreting relatively efficiently because its
\ion{H}{2} region is still confined (see below), thus resulting in a high 
accretion luminosity. However, in this latter case we do not expect
accretion to be significantly affected by radiation pressure because
much of the luminosity will be directed in polar directions and most
of the accretion is of neutral gas from the equatorial regions.

\begin{figure}
  \includegraphics[height=0.3\textheight]{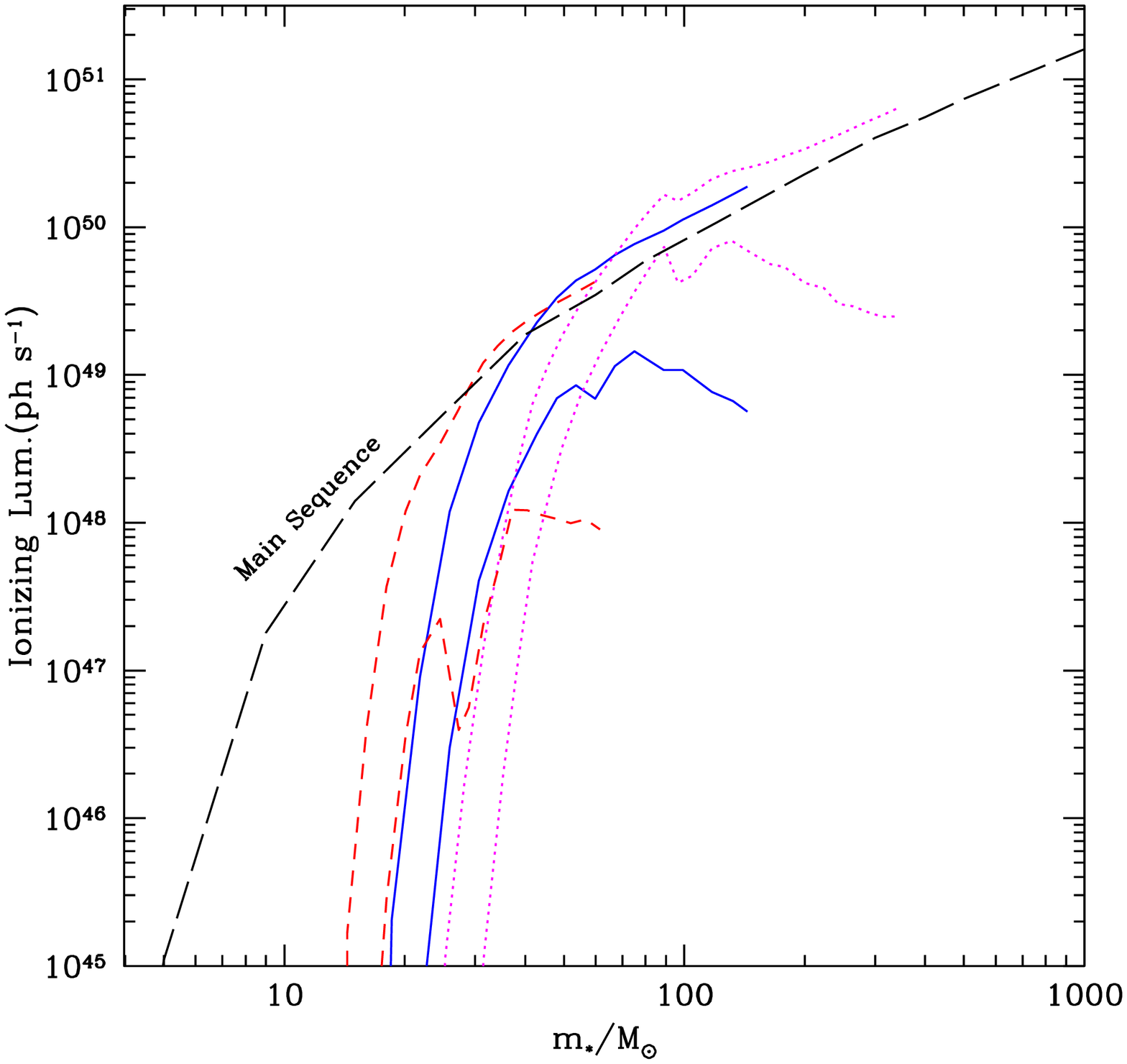}
\caption{
  \label{fig:Sstarreview} Protostellar H-ionizing photon luminosities
for models with $K^\prime=0.5,1,2$ ({\it dashed, solid, dotted} lines). The
total luminosity is shown with the higher line and the accretion
luminosity (boundary layer $+$ inner disk) is shown with the lower
line. The zero age main sequence H-ionizing luminosities
\cite{schaerer2002} are indicated with the {\it long-dashed} line.  }
\end{figure}

As discussed in McKee \& Tan \cite{mckee2008}, we expect ionizing
feedback to be more important than continuum radiation pressure at
reducing accretion efficiency and eventually setting the final stellar
mass. The H-ionizing photon luminosities for the $K^\prime=0.5,1,2$
models are shown in Fig.~\ref{fig:Sstarreview}. At lower masses, the
protostars are much larger than the corresponding main sequence stars
and so have cooler photospheric temperatures and smaller ionizing
luminosities. Once the protostar is older than about its instantaneous
Kelvin-Helmholz time, which occurs at higher masses for higher
accretion rates, the star contracts towards the main sequence and the
ionizing luminosity rises accordingly.

\begin{figure}
  \includegraphics[height=0.7\textheight]{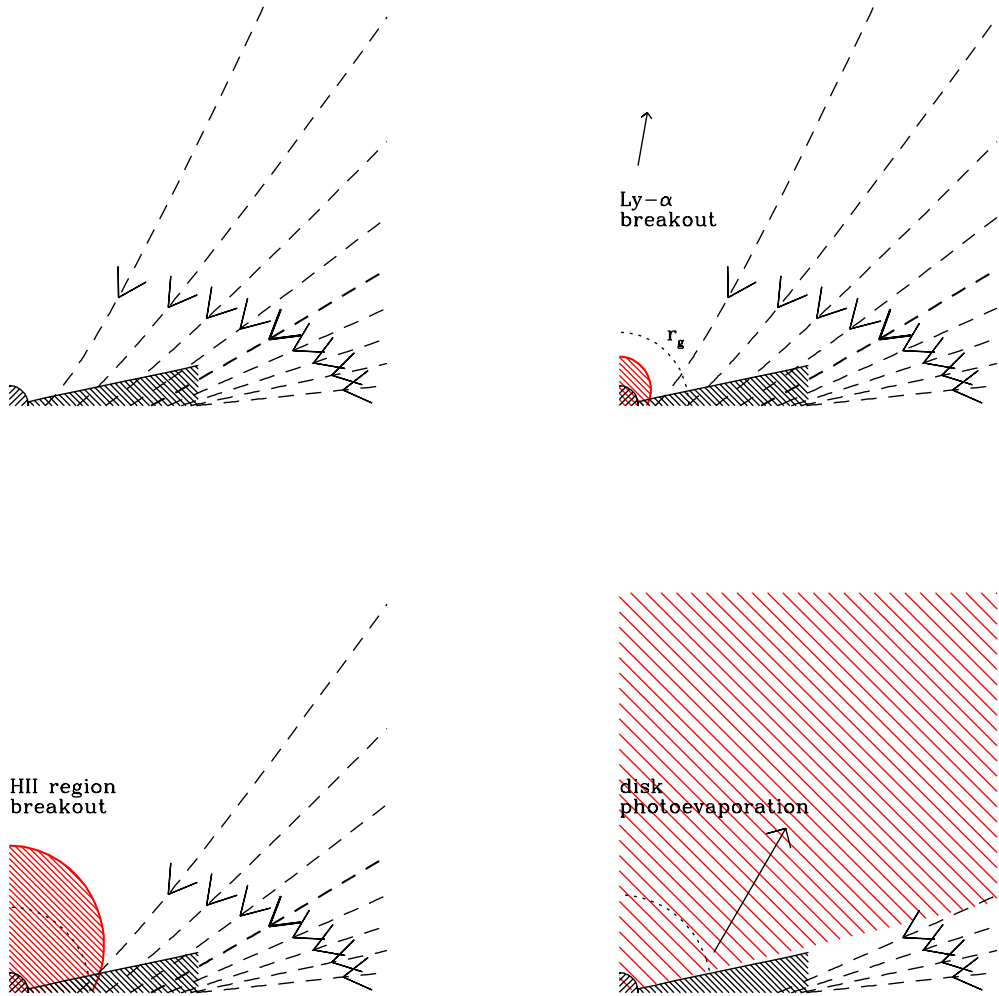}
\caption{\label{fig:overview}Overview of the accretion geometry and
feedback processes involved in primordial star formation. (a) {\it Top left:}
Cross section of the accretion geometry: the dashed lines show
streamlines of the rotating, infalling gas, with figure of revolution
from each streamline separating 10\% of the total infall from this
hemisphere. The aspect ratio of the accretion disk is realistic, while
the size of the star has been exaggerated for clarity. (b) {\it Top right:}
The red shaded region shows the extent of the
\ion{H}{2} region, which at this relatively early stage is still
confined inside the gravitational radius for the escape of ionized
gas, $r_g$. \lal\ radiation pressure feedback should be strong enough
to prevent accretion in the polar directions. (c) {\it Bottom left:} The
stellar mass and ionizing luminosity have grown, and the
\ion{H}{2} region is just in the process of breaking out of the
accretion flow. Once a significant volume beyond $r_g$ is ionized,
accretion from these directions is expected to be shut off. (d) {\it Bottom
right:} Final stage of accretion involves shadowing of the equatorial
region by the disk, which at the same time is photoevaporated. The
competition between this photoevaporative outflow and the residual
accretion rate sets the final mass of the star.}
\end{figure}

Figure \ref{fig:overview} gives an overview of the ionizing feedback
processes occurring near the protostar.  At early stages the ionizing
flux from the protostar is smaller than the flux of accreting neutral
atoms to its surface and the \ion{H}{2} region is confined there. As
the ionizing luminosity increases, the \ion{H}{2} region can begin to
expand into the infalling envelope, penetrating furthest along the
rotation axes. At the ionized-neutral boundary, radiation pressure
feedback is exerted due to resonant scattering of FUV radiation in the
\lal\ damping wings. As a result of the high column densities of
neutral gas around the \ion{H}{2} region, this radiation is trapped
and the pressure amplified by large factors. For typical rotation
parameters $f_{\rm Kep}\simeq0.5$, this radiation pressure becomes
larger than the ram pressure of the infalling gas in the polar
directions for stellar masses of order 20$\sm$. However, once the
infall is reversed at the poles, the \lal\ photons can escape and the
accretion in other directions proceeds relatively unimpeded. Thus we
expect \lal\ radiation pressure to have only a minor effect on the
accretion efficiency.

The next feedback effect to occur is the expansion of the \ion{H}{2}
region to distances larger than the gravitational escape radius $r_g$
for ionized gas. This distance is
\beq
\label{eq:rg}
r_g  \equiv \frac{G\phiedd\msd}{c_i^2}= 260\phiedd\left(\frac{2.5}{T_4}\right)\msdt~{\rm AU},
\eeq
where $T_4$ is the ionized gas temperature in units of $10^4$~K and we
have taken the gravitating mass to be $\msd$ and we have allowed for
the decrease in the radius due to radiation pressure from electron
scattering through the factor
\beq
\phiedd\equiv 1-\frac{L}{L_{\rm Edd}}\; ,
\label{eq:phiedd}
\eeq
where $\ledd=
4\pi G m c/\kappa_{\rm Thomson}$ is the Eddington limit. 

\begin{figure}
  \includegraphics[height=0.3\textheight]{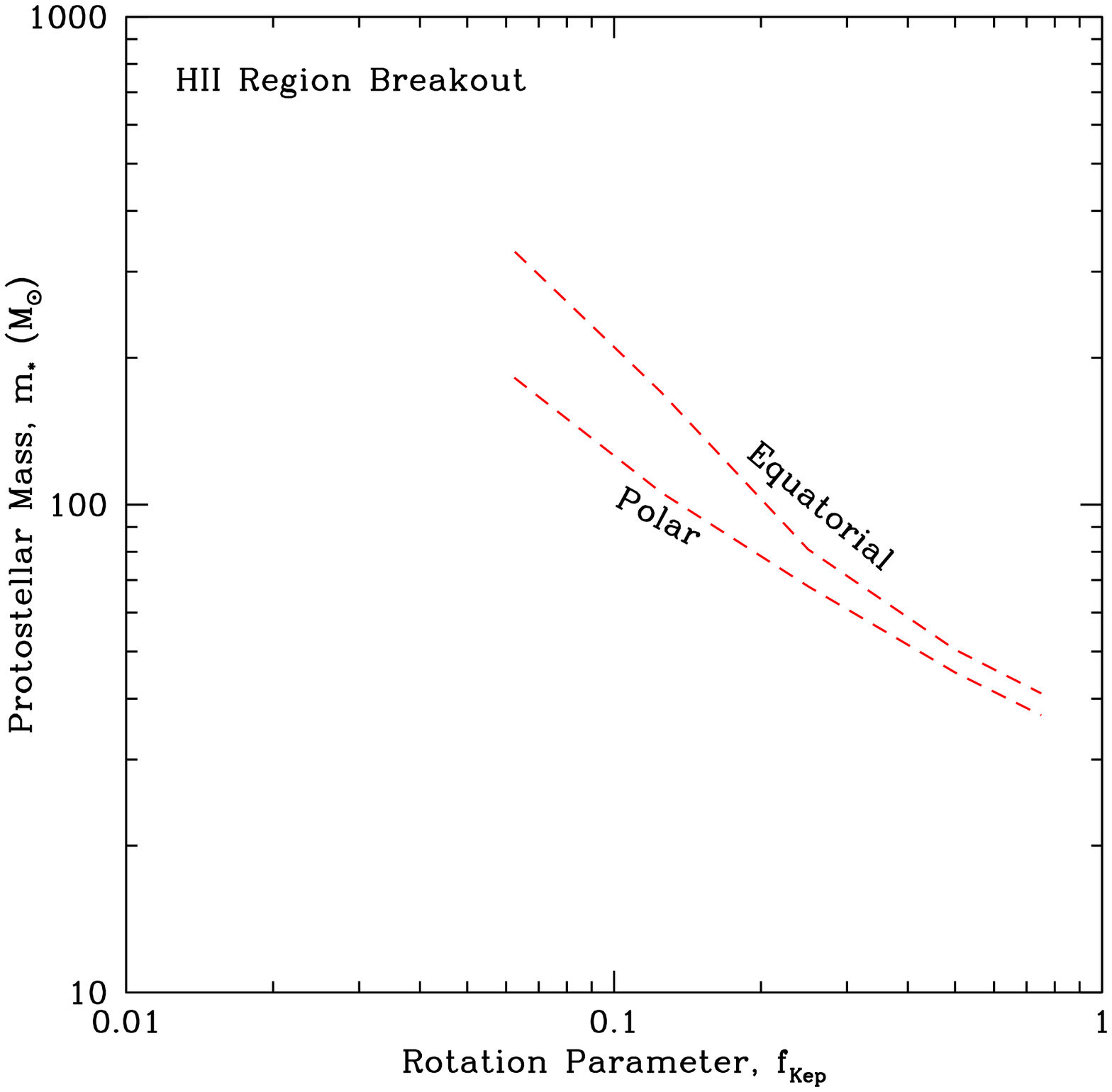}
\caption{
  \label{fig:hiibreakout}
Mass scales of \ion{H}{2} region break out as a function of the
rotation parameter $f_{\rm Kep}$. The {\it lower dashed} line marked
``Polar'' shows the mass scale of the protostar at which the
\ion{H}{2} region reaches $r_g$ (based on star plus disk mass) along
the rotation axis of the protostar. The {\it upper dashed} line marked
``Equatorial'' shows the mass scale of the protostar when the
\ion{H}{2} region reaches $r_g$ in a direction just above the disk
plane ($0.9\pi/2$ from the rotation axis).
}
\end{figure}

Shortly after the \ion{H}{2} region reaches $r_g$, pressure forces in
the ionized gas become large enough to reverse the infall to the
protostar. This effect will occur first in the polar directions,
typically at about $50\sm$ in our fiducial model. For rotation
parameters $f_{\rm Kep}\simeq 0.5$, expansion in the equatorial
directions (just above the accretion disk) occurs very soon after
polar breakout (see Fig.~\ref{fig:hiibreakout} and Table
\ref{tab:m}). Higher accretion rates and smaller rotation parameters 
result in larger mass scales for \ion{H}{2} region breakout.

\begin{table}
\begin{tabular}{cccccc}
\hline
 \tablehead{1}{r}{b}{$K^\prime$}
  & \tablehead{1}{r}{b}{$f_{\rm Kep}$}
  & \tablehead{1}{r}{b}{$T_i/(10^4\:{\rm K})$}
  & \tablehead{1}{r}{b}{$m_{\rm *,pb}$ ($\sm$)\tablenote{Mass scale of HII region polar breakout.}}
  & \tablehead{1}{r}{b}{$m_{\rm *,eb}$ ($\sm$)\tablenote{Mass scale of HII region near-equatorial breakout.}}
  & \tablehead{1}{r}{b}{$m_{\rm *,evap}$ ($\sm$)\tablenote{Mass scale of disk photoevaporation limited accretion.}}\\
\hline
1 & 0.5 & 2.5 & 45.3 & 50.4 & 137\tablenote{Fiducial model.}\\
\hline
1 & 0.75 & 2.5 & 37 & 41 & 137\\
1 & 0.25 & 2.5 & 68 & 81 & 143\\
1 & 0.125 & 2.5 & 106 & 170 & 173\\
1 & 0.0626 & 2.5 & 182 & 330\tablenote{This mass is greater than $m_{\rm *,evap}$ in this case because it is calculated without allowing for a reduction in $\dot{m}_*$ during the evolution due to polar HII region breakout.} & 256\\
\hline
1 & 0.5 & 5.0 & 35 & 38 & 120\\
1 & 0.25 & 5.0 & 53.0 & 61 & 125\\
\hline
0.5 & 0.5 & 2.5 & 23.0 & 24.5 & 57\\
\hline
2.0 & 0.5 & 2.5 & 85 & 87 & 321\\
\hline
\end{tabular}
\caption{Mass Scales of Population III.1 Protostellar Feedback}
\label{tab:m}
\end{table}

\begin{figure}
  \includegraphics[height=0.3\textheight]{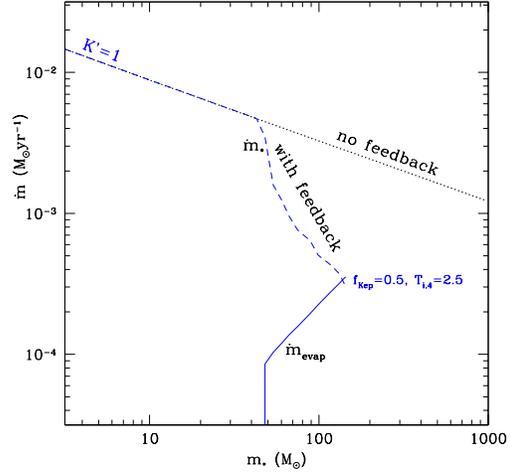}
\caption{
  \label{fig:theend1}
Feedback-limited accretion: fiducial case. The evolution of the
accretion rate versus protostellar mass is shown for the fiducial
model ($f_{\rm Kep}=0.5$, $K^\prime =1$, $T_i = 25,000$~K) in the
cases of ``no feedback'' and ``with feedback''. In the latter, the
accretion efficiency is reduced as the \ion{H}{2} region expands to
$r_g$ and beyond. However, accretion is allowed to continue from
directions that are shadowed by the disk photosphere. The disk
structure and protostellar structure and feedback are calculated
self-consistently given the evolution in $\dot{m}_*$. Also shown is
the photoevaporative mass loss rate, $\dot{m}_{\rm evap}$, which
starts once the \ion{H}{2} region has broken out in the equatorial
direction and grows as the ionizing flux increases. We see that this
mass loss rate becomes greater than the accretion rate at $m_*\simeq
137\:M_\odot$, and we identify this mass scale as our best estimate of
initial mass scale of the first stars.
}
\end{figure}

The protostar and accretion envelope should evolve to a state in which
the entire region above and below the accretion disk has been ionized
(Fig.~\ref{fig:overview}d). No accretion will occur from
these directions. The accretion disk, which is still mostly neutral in
its outer regions, has a finite thickness and will shield an extended
equatorial region from direct ionizing flux from the protostar,
allowing accretion to be able to continue from these regions. The
reduction in accretion rate caused by the expansion of the \ion{H}{2}
region and the fraction allowed in the disk-shadowed region is shown
in Fig.~\ref{fig:theend1}, where the fiducial no-feedback case is also
shown.

\begin{figure}
  \includegraphics[height=0.3\textheight]{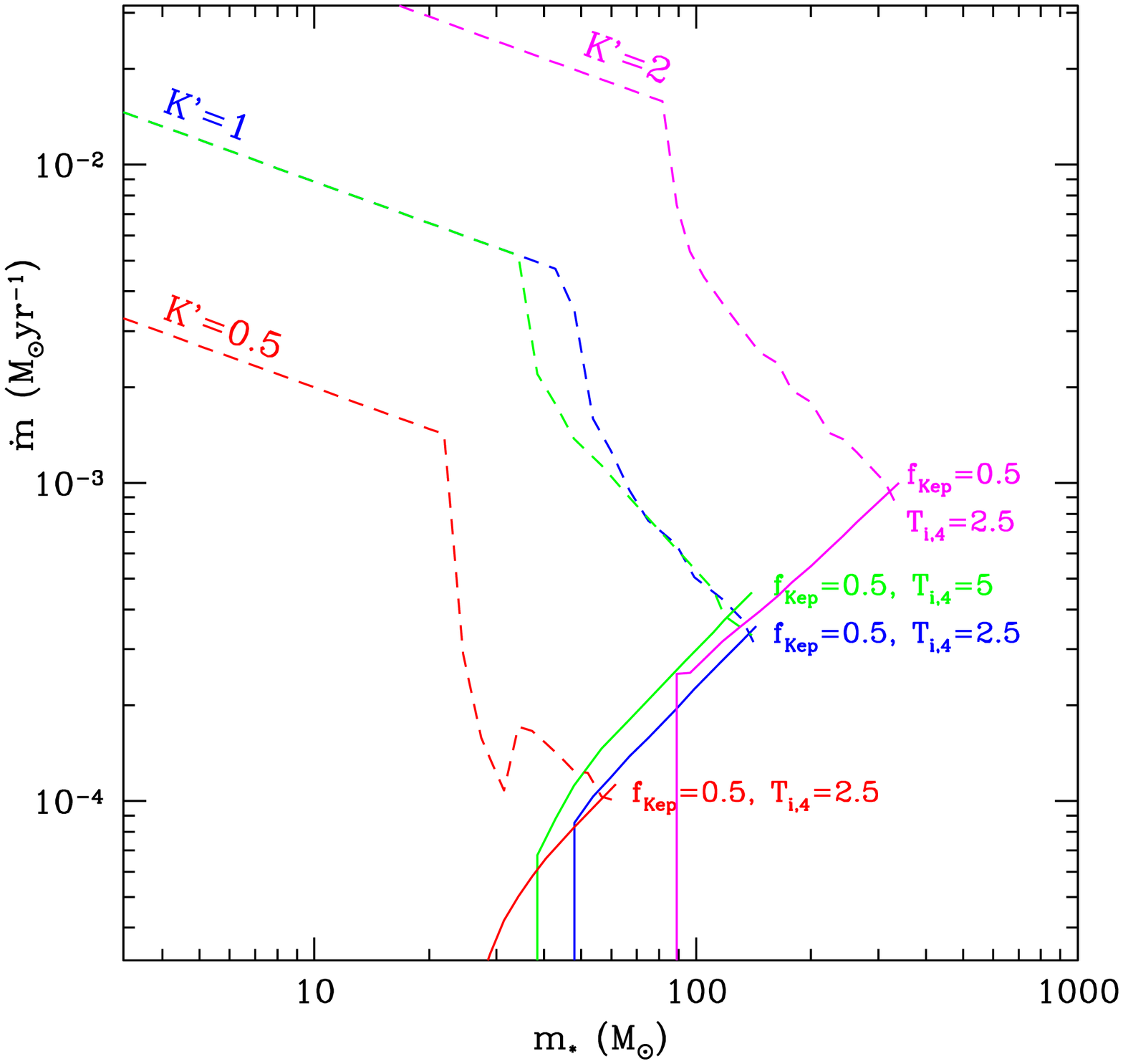}
\caption{
  \label{fig:theend2}
Feedback-limited accretion: effect of ionized gas temperature and
accretion rate. The fiducial model ($f_{\rm Kep}=0.5$, $K^\prime =1$,
$T_{i,4} = 2.5$~K) shown in Fig.~\ref{fig:theend1} is compared to
models in which one parameter has been changed: a model with
$T_{i,4}=5$ and two models with $K^\prime=0.5,2$. The dashed lines
show the accretion rate to the star, $\dot{m}_*$, and the solid lines
show the photoevaporative mass loss rate, $\dot{m}_{\rm evap}$. The
change in temperature causes relatively minor differences, while the
change in $K^\prime$, equivalent to a change in $\dot{m}_*$ of factors
of 4.4 above and below the fiducial level, leads to roughly a factor
of 2.4 change in the final stellar mass. Note the increase in
$\dot{m}_*$ for the $K^\prime=0.5$ case at around $35\sm$ is due to a
thickening of the inner accretion disk as the star contracts down to
its main sequence configuration and assumes material at large
distances still remains to be accreted in the enlarged shadowed
region.
}
\end{figure}

Also shown in Fig.~\ref{fig:theend1} is the mass loss rate due to the
process of disk photoevaporation \cite{hollenbach1994}. Ionization
from the protostar creates an ionized atmosphere above the neutral
accretion disk, which then scatters some ionizing photons down on to
the shielded region of the outer disk, beyond $r_g$. An ionized
outflow is driven from these regions at a rate
\beq
\dot{m}_{\rm evap} 
\simeq 4.1\times 10^{-5} S_{\rm 49}^{1/2}T_4^{0.4}\msdt^{1/2}~~\smyr,
\eeq
where $S_{\rm 49}$ is the H-ionizing photon luminosity in units of
$10^{49}$ photons~$\rm s^{-1}$. When this rate becomes comparable to
the accretion rate to the disk, then we expect further growth of the
stellar mass to be very limited. For simplicity we equate the final
stellar mass to that mass at which the photoevaporative mass loss rate
equals the accretion rate. It is about $140 M_\odot$ in the fiducial
case, and Table~\ref{tab:m} and Fig.~\ref{fig:theend2} summarize other
cases.

An analytic estimate can be made for the final protostellar
mass. Assuming the H-ionizing photon luminosity is mostly due to the
main sequence luminosity of the star we have
\beq
S\simeq 7.9\times 10^{49}\; \phi_S \mst^{1.5}~~~~{\rm ph\ s^{-1}},
\label{eq:ss}
\eeq which for $\phi_S=1$ is a fit to Schaerer's (2002) results for
the ionizing luminosity of main sequence primordial stars; the fit is
accurate to within about 5\% for $60 M_\odot \lesssim m_*\lesssim 300 M_\odot$.
Then the photoevaporation rate becomes
\beqa
\dot{m}_{\rm evap} & = & 1.70\times 10^{-4}\phi_S^{1/2}(1+f_d)^{1/2}\nonumber\\
 &  \times & \left(\frac{T_4}{2.5}\right)^{0.4} \mst^{5/4}~~\smyr.
\label{eq:evap2}
\eeqa
The accretion rate onto the star-disk system is given by equation
(\ref{eq:mdot}). Equating this with equation (\ref{eq:evap2}), we find
that the resulting maximum stellar mass is
\beq
\label{eq:maxevap}
{\rm Max}\; m_{*f,2}=
6.3\;\frac{\esd^{28/47}\esdb^{12/47}
\phi_S^{14/47}K'^{60/47}}{(1+f_d)^{26/47}} 
        \left(\frac{2.5}{T_4}\right)^{0.24}\; .
\eeq
Recall that $\esd$ is the instantaneous star formation
efficiency---i.e., the ratio of the accretion rate onto the star to
the rate that would have occurred in the absence of feedback. Here
this ratio is just the shadowing factor, $\fsh$, i.e. the fraction of
the sky as seen from the protostar that is blocked by the disk. For
this simple analytic estimate, we make the approximation that the
shadowing sets in when the stellar mass reaches $m_1$, so that
$\esd=1$ until the mass of the central star reaches $m_1$ and
$\esd=\fsh$ thereafter. It is then straightforward to show that
\beq
\esdb=\frac{\fsh}{1-(1-\fsh)(m_1/\msd)},
\eeq
provided that $\msd\geq m_1$. Note that the average efficiency $\esdb=1$
at the onset of shadowing $(\msd=m_1)$ and that $\esdb\rightarrow\fsh$
at late times $\msd\gg m_1$. Normalizing $\fsh$ to a typical value of 0.2
we find
\beqa
\label{eq:maxm}
{\rm Max}\; m_{*f,2}& = & 1.45\, K'^{60/47}  \left(\frac{2.5}{T_4}\right)^{0.24}\nonumber\\
 & \times &  \left(\frac{\fsh}{0.2}\right)^
        {28/47}\left(\frac{\esdb}{0.25}\right)^{12/47},
\eeqa
where we have set the ionizing luminosity factor $\phi_S=1$ and the
disk mass fraction $f_d=\frac 13$; we have normalized $\esdb$ to a
value of 0.25, which is approximately correct for $K'=1$ and for
$m_1\simeq 50 M_\odot$ and $\msd=200\sm$. This analytic estimate
therefore also suggests that for the fiducial case ($K'=1$) the mass
of a Pop III.1 star should be $\simeq 140 M_\odot$.

The uncertainties in these mass estimates include: (1) the assumption
that the gas distribution far from the star is approximately spherical
--- in reality it is likely to be flattened towards the equatorial
plane, thus increasing the fraction of gas that is shadowed by the
disk and raising the final protostellar mass; (2) uncertainties in the
disk photoevaporation mass loss rate due to corrections to the
Hollenbach et al. \cite{hollenbach1994} rate from the flow starting
inside $r_g$ and from radiation pressure corrections; (3)
uncertainties in the \ion{H}{2} region breakout mass due to
hydrodynamic instabilities and 3D geometry effects; (4) uncertainties
in the accretion rate at late times, where self-similarity may break
down \cite{bromm2004}; (5) the effect of rotation on protostellar
models, which will lead to cooler equatorial surface temperatures and
thus a reduced ionizing flux in the direction of the disk; (6) the
simplified condition, $\dot{m}_{\rm evap}>\mdsd$, used to mark the end
of accretion; and (7) the possible effect of protostellar outflows
(discussed below).

\subsubsection{Mechanical Feedback}

Protostellar outflows, thought to be launched by large-scale magnetic
fields threading the inner accretion disk, are ubiquitous from
present-day protostars. The momentum flux in these flows typically
exceeds that due to radiation pressure by several orders of
magnitude. As described above, Tan \& Blackman \cite{tan2004b}
discussed how turbulence in a stratified disk that generates helicity
is the necessary condition for production of dynamically-strong
large-scale magnetic fields by dynamo action. They then considered the
effect of such an outflow on the surrounding minihalo gas, following
the analysis of Matzner \& McKee \cite{matzner2000}. The force
distribution of centrifugally-launched hydromagnetic outflows is
collimated along the rotation axes, but includes a significant
wider-angled component. Matzner \& McKee \cite{matzner1999} showed
that far from the star the angular distribution of the momentum in a
radial hydromagnetic wind can be approximated by 
\beq
\frac{dp_w}{d\Omega}
= \frac{p_w}{4\pi} \frac{1}{{\rm ln}(2/\theta_0)(1+\theta_0^2 -
\mu^2)},
\label{pangular}
\eeq where $\mu={\rm cos}\theta$ and $\theta_0$ is a small angle,
which is estimated to be $\sim 0.01$ for winds from low-mass stars.
The total momentum of the outflow, $p_w$, is calculated by assuming a
fixed ratio of mass injected into the outflow relative to mass
accretion to the star, $\dot{m}_w/\dot{m}_*=0.084$ in the fiducial
model of Tan \& Blackman. The speed of the outflow is set equal to the
escape speed from the stellar surface. As the star grows in mass and
contracts to the main sequence the cumulative specific momentum of the
outflow per stellar mass rises from $p_w/m_* \simeq 7\:{\rm
km\:s^{-1}}$ when $m_*\simeq10\sm$ to $p_w/m_* \simeq 100\:{\rm
km\:s^{-1}}$ when $m_*\simeq100\sm$.

To evaluate the efficiency of star formation from a core we find the
angle, $\theta_{\rm esc}\equiv{\rm cos}^{-1}\mu_{\rm esc}$, where the
wind sweeps up core material to the surface escape speed of the core,
$v_{\rm esc,c}$. The location of the surface of Pop III.1 cores is not
particularly well defined: the fiducial core of Tan \& McKee
\cite{tan2004}, based on the simulation results of Abel, Bryan, \&
Norman \cite{abn2002}, has about 1000 $\sm$ of virialized baryonic
material with a few $\times 10^4 \sm$ of surrounding gas still
infalling. These cores have $v_{\rm esc,c}=3.22 (M/1000M_\odot)^{-1/7}
K'^{5/7}\:{\rm km/s}$.  To find the speed of the swept-up gas, the
analysis assumes thin, radiative shocks, purely radial motion and
monopole gravity. We consider cores with angular mass distributions of
the form $dM/d\Omega =(1/4\pi) Q(\mu) M$ with $Q(\mu)=(1-\mu^2)^n /
\int_0^1 (1-\mu^2)^n \:d\mu$, and evaluate models with $n$, ranging
from $0$, isotropic, to $4$, which describes a flattened distribution
that mimics the effect of some rotational support. In the notation of
Matzner \& McKee \cite{matzner2000}, the escape condition is given by
$(1+\theta_0^2 - \mu_{\rm esc}^2) Q(\mu_{\rm esc}) = m_*/(X M)$, where
$X=0.132 c_g [{\rm ln}(2/\theta_0) / {\rm ln} 200] v_{\rm esc,c,5} [
(p_w/m_*)/ 40{\rm km\:s^{-1}} ]^{-1}$, and $c_g$ is a factor of order
unity that accounts for the effects of gravity on shock propagation.
We estimate $c_g\simeq4.6$, for steady winds
that decouple from the swept-up shell at the core edge in a core with
$k_\rho = 2.2$ (which is the fiducial value of Tan \& McKee
\cite{tan2004}). We ignore the influence
of material beyond the core ``edge'', which is partly balanced by our
neglect of the wind's influence on the shell beyond this point.

\begin{figure}
  \includegraphics[height=0.3\textheight]{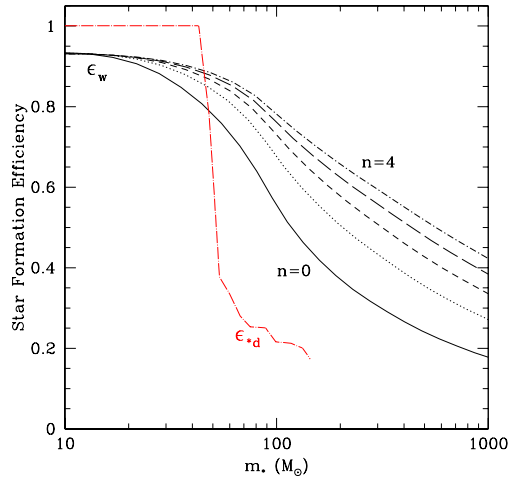}
\caption{
  \label{fig:ecore} Evolution of star formation efficiency,
$\epsilon_w$, due to erosion of a 1000$\sm$ gas core by protostellar outflows
winds for the fiducial $f_{\rm Kep}=0.5$ and $K'=1$ case, with
$\alpha_{\rm ss}=0.01$, and ignoring radiative feedback processes
\cite{tan2004b}. The density profile of the initial core is specified
by $dM/d\Omega =(1/4\pi) Q(\mu) M$, with $\mu = {\rm cos}
\theta$ and $Q(\mu)=(1-\mu^2)^n / \int_0^1 (1-\mu^2)^n \:d\mu$. {\it Solid}
line is $n=0$ (isotropic core), {\it dotted} is $n=1$, {\it dashed} is
$n=2$, {\it long-dashed} is $n=3$, {\it dot-dashed} is $n=4$. The {\it
dot-long-dashed} line shows the instantaneous efficiency, $\esd$, due
to radiative feedback processes (i.e. \ion{H}{2} region breakout,
limited by disk-shadowing) for the fiducial model ($f_{\rm Kep}=0.5$,
$K'=1$, $T_{i,4}=2.5$), indicating their greater importance at
$m_*\gtrsim 50\sm$.}
\end{figure}

The star formation efficiency due to protostellar outflow winds is
given by \beq \epsilon_w =
\frac{1}{1+(\dot{m}_w/\dot{m}_*)(1-\phi_w)} \int_0^{\mu_{\rm esc}} Q
\:{\rm d}\mu,
\label{eq:ecore}
\eeq where $\phi_w\equiv \int_0^{\mu_{\rm esc}} P\:{\rm d}\mu$ and
$P(\mu)$ is the dimensionless force distribution in angle.  Note that
this definition of efficiency includes the effect of diversion of a
fraction $\dot{m}_w/\dot{m}_* = 0.084$ of the stellar accretion rate
into outflow material.  Equation \ref{eq:ecore} can be evaluated as a
function of $m_*$ (see Fig.~\ref{fig:ecore}). $\theta_{\rm esc}$
gradually increases and $\epsilon_w$ decreases as the star grows and
injects more momentum into its surroundings. However, note that
$\epsilon_w$ is an upper limit to the star formation efficiency if it
is evaluated before the final stellar mass is reached since some
material at $\theta>\theta_{\rm esc}$ would later be ejected. Note
also that the simple models shown here do not self-consistently
account for the effects that a smaller mass accretion rate would have
on protostellar evolution and outflow strength. Nevertheless they show
that by the time $m_*$ has reached a hundred to a few hundred solar
masses, the outflows have been able to eject about half the initial
core material.

Figure~\ref{fig:ecore} also shows the instantaneous accretion
efficiency to the star$+$disk, $\epsilon_{*d}$, due to radiative
feedback and disk-shadowing for the fiducial $K^\prime=1$, $f_{\rm
Kep}=0.5$, $T_{i,4}=2.5$ model. Comparison of these efficiencies is
complicated by the fact that, as mentioned, for $\epsilon_w$ at a
given $m_*$, some gas at angles $>\theta_{\rm esc}$ will eventually be
ejected if the outflow continues. Nevertheless, we see that radiative
feedback and disk-shadowing are more important, i.e. lead to smaller
efficiencies, than mechanical feedback for masses $\gtrsim 50\sm$ in
the fiducial case. The outflow would need to operate until $m_*\sim
1000\sm$ to reduce $\epsilon_w$ to the value of $\epsilon_{*d}$ seen
at $m_*\simeq 100\sm$. The possibility of lateral deflection of gas
streamlines by the outflow is not treated by the above analysis, but
we expect this would tend to increased $\epsilon_w$. This effect can
be studied with numerical simulations. Machida et
al. \cite{machida2006} have carried out 3D MHD simulations of the very
earliest stages of protostellar outflow feedback, up to times when the
protostar has just formed with $m_*\lesssim 0.01\sm$. Such simulations
need to be advanced to later stages to explore the effects of a
realistic 3D geometry on the outflow-core interaction and star
formation efficiency.

While mechanical feedback is unlikely to dominate over radiative
feedback, it may influence details of the radiative feedback
processes. Even a small outflow cavity when $m_*\sim 20\sm$ will help
dissipate Ly-$\alpha$ photons and prevent a build up of dynamically
important radiation pressure at this stage. An outflow will tend to
facilitate \ion{H}{2} region breakout in polar directions by lowering
gas densities and injecting additional momentum. As it is launched
from the surface of the disk, a dense bipolar outflow can confine the
protostellar ionizing flux in equatorial directions and prevent it
reaching outer disk to cause photoevaporation
\cite{tan2004c}. However, Tan \& Blackman \cite{tan2004b} showed that the 
fiducial outflow would not be dense enough to significantly shield the
disk from the ionizing luminosities of primordial protostars once
$m_*\gtrsim40\sm$. On the other hand, the outflow is likely to impact
the structure of the ionized disk atmosphere and thus affect the
photoevaporation mass loss rate, perhaps similar to the ``strong
stellar wind'' case of Hollenbach et
al. \cite{hollenbach1994}.

\subsection{Impact of Pop III.1 Stars for Larger-Scale Feedback and Metal Production}

The predicted final protostellar masses (i.e. the initial stellar
masses) for the range of model parameters we have considered are
summarized in Table~\ref{tab:m}. The stars are all ``massive''
(i.e. $m_*>8\sm$) and will thus have significant radiative and
mechanical feedback on their surroundings, and the potential for
substantial metal enrichment via core-collapse or pair-instability
supernovae.

Ionization, outflow, wind and supernova feedback can lead to the
disruption and unbinding of gas from nearby minihalos
In particular ionization is likely to be the first and
most important feedback effect to initiate unbinding: ionizing
radiation heats up the gas to temperatures $\sim 25,000$~K
\cite{giroux1996,shapiro2004}, depending on the temperature of the
radiation field, with corresponding sound speeds that are much larger
than the few km/s typical of minihalos. Note, however, that in
addition to unbinding gas from minihalos, the pressure forces
associated with ionized regions can shock-compress adjacent neutral
regions and promote gravitational collapse, and this possibility is
discussed in the next section. Stars with masses in the range shown in
Table~\ref{tab:m} are able to completely ionize at least their local
minihalo, and more typically a much large volume of the surrounding
intergalactic medium. For recent numerical simulations of
photoevaporation of minihalos see Whalen et al. \cite{whalen2008} and
references therein.

Lyman-Werner band radiation destroys \htwo\ molecules, reducing the
cooling efficiency of the gas.  We can estimate the distance over
which star formation is suppressed from the work of Glover \& Brand
\cite{glover2001}. Assuming that the core is in approximate
hydrostatic equilibrium and is characterized by an entropy parameter
$K$, we find \cite{mckee2008} that the time to dissociate \htwo\ is
less than the free-fall time $\tff$ if the core is within a distance
\beq
D=\left(\frac{S_{\rm LW}}{10^{49}\ \s^{-1}}
        \frac{10^{-3}}{x_2}
        \frac{f_{\rm abs}f_{\rm diss}}{0.01}\right)^{1/2}
        \frac{24}{\bar n_4^{21/40}K'^{1/4}}{\rm pc},
\eeq
of the protostar, where $S_{\rm LW}$ is the photon luminosity in the
Lyman-Werner bands, $x_2$ is the fractional abundance of \htwo,
$f_{\rm abs}$ is the fraction of the Lyman-Werner flux absorbed by the
\htwo, $f_{\rm diss}$ is the fraction of absorptions that result in
dissociation, and $\bar n$ is the mean density of H nuclei.  Thus,
even a $100 \sm$ star, which has $S_{\rm LW}\simeq 10^{49}$~s\e, can
suppress star formation in an existing core only if the core is
relatively nearby.  A more detailed analysis by Susa \cite{susa2007}
comes to similar conclusions (see also Ahn \& Shapiro \cite{ahn2007}
and Whalen et al. \cite{whalen2008}).

If gas ionized by the Pop III.1 star has a chance to recombine,
e.g. once the star completes its nuclear burning, the relatively high
residual electron fraction catalyzes molecule formation, particularly
of HD, which can dramatically enhance subsequent cooling
\cite{uehara2000}. This is the most likely scenario for the next 
generation of star formation, Population III.2 (discussed in the next
section).

Lacking metal lines in their atmospheres, main-sequence and
post-main-sequence Pop III stars are expected to have quite weak
stellar winds \cite{kudritzki2002}. This also implies their mass at
the end of stellar evolution should be similar to their initial
mass. Rotation is expected to enhance mass loss
\cite{meynet2006}, although the final mass is still thought to be a 
significant fraction of the initial mass.

Thus the range of final protostellar masses predicted by models of
radiative and mechanical feedback (Table~\ref{tab:m}) imply a range of
final pre-core-collapse stellar masses that potentially overlap the
range of masses necessary to produce pair instability supernovae,
$140-260\;M_\odot$ in the models of Heger \& Woosley \cite{heger2002}.
Rotation may lower these limits (S. Woosley, private comm.). The lack
of the expected nucleosynthetic signature of such supernovae in the
abundance patterns of very metal poor halo stars
\cite{tumlinson2004}, could indicate that such massive Pop III.1 stars
were relatively rare or that they tended to enrich regions not probed
by typical halo stars, perhaps the centers of larger galactic halos.
The conclusion by Scannapieco et al. \cite{scannapieco2006} that Pop
III star formation should be fairly widespread in regions now probed
by Galactic halo stars, can be reconciled with the abundance pattern
observations if most of this star formation leads to either Pop III.1
stars from relatively low entropy ($K^\prime \lesssim 1$) gas cores or
Pop III.2 stars (discussed below) that also have a mass scale below
the pair instability threshold (see also the study by Greif \& Bromm
\cite{greif2006}). Further work is
required to determine the range of pre-stellar core parameters,
primarily $K^\prime$ and $f_{\rm Kep}$, exhibited in cosmological
simulations, in order to predict the frequency of pair instability
supernovae.

\section{Population III.2 Star Formation}

Population III.2 stars are those with near primordial composition, so
that their formation and stellar evolution are independent of
metallicity, but with initial conditions ``significantly'' affected by
radiative and/or mechanical feedback from previous stellar generations
(that could be Pop III.1, Pop III.2, or even Pop II or Pop I) or AGN. 

We expect radiative feedback (especially FUV destruction of \htwo ) to
be far more pervasive than the direct mechanical feedback associated
with protostellar outflows, winds or supernovae, although shocks
driven by D-type ionization fronts will be even more penetrating than
ionization into dense regions. Such shock compression may be important
in promoting star formation in neighboring minihalos of intermediate
central density, $10\:{\rm cm^{-3}}\lesssim n_{\rm H}
\lesssim 1000\:{\rm cm^{-3}}$ \cite{whalen2008}, although the properties 
of the stars that subsequently form under these conditions remain to
be determined. Stacy \& Bromm
\cite{stacy2007} have discussed the impact of cosmic rays produced
from Pop III stars, which would also be highly penetrating, although
estimates of their production rate are uncertain.

In gas that is no longer exposed to significant ionizing and
dissociating flux and that has a chance to recombine, the relatively
high residual electron fraction catalyzes molecule formation,
particularly of HD, which can dramatically enhance subsequent cooling,
perhaps reducing the characteristic star-formation
mass\cite{uehara2000, greif2006}. Such a reduction is in qualitative
agreement with our models with lower values of $K^\prime$, which is
proportional to the temperature (eq. \ref{eq:kp}): such stars reach
the main sequence (and associated higher ionizing luminosities) at
lower protostellar masses. They also have smaller rates of accretion
that they need to disrupt via disk photoevaporation.  The precise mass
range that is applicable for typical Pop III.2 stars requires more
detailed numerical and analytic study, but initial estimates, based on
the $K^\prime=0.5$ model, suggest that their masses are significantly
smaller than those necessary to produce pair instability supernovae
\cite{heger2002}.

\section{The Transition to Population II}

Metal atoms and dust grains provide additional mechanisms for the gas
to cool. For a given temperature, density and ionization history there
is some ``critical metallicity'' at which these will start to
influence star formation by dominating over \htwo\ and HD cooling.

The value of this critical metallicity for star formation is
uncertain, with estimates ranging from $\sim 10^{-6}Z_\odot$ if the
cooling is dominated by small dust grains that contain a significant
fraction of the metals \cite{omukai2005} to $\sim 10^{-3.5}Z_\odot$ if
the cooling is dominated by the fine structure lines of C and O and
there is negligible \htwo\ \cite{bromm2003}; if \htwo\ cooling is
included, Jappsen et al. \cite{jappsen2007} argue that there is no
critical metallicity for gas-phase metals.

A detailed understanding of the transition to Pop II star formation
will require knowledge of the Pop III.1 and III.2 IMFs, their mass
loss during stellar evolution due to stellar winds and the metal and
dust content of these winds, the metal and dust content of their
supernova ejecta, the mixing of wind and supernova ejected metals
and dust into the surrounding (dense) gas that subsequently becomes
gravitational unstable. It is fair to say that such a complicated
problem, with so many uncertain inputs, remains far from solved. The
general consensus is that, with additional coolants the gas can
fragment into smaller pre-stellar cores that will tend to have smaller
accretion rates. Both of these effects should lead to an IMF peaked at
smaller masses than that of Pop III stars.

\section{Comparison to Present-Day Star and Star Cluster Formation}

\subsection{Initial Conditions}

The mean gas densities and velocity dispersions of Pop III.1 minihalos
and present-day star-forming gas clumps \cite{mueller2002} are quite
similar. The major difference is that in present-day gas, cooling by
molecules and dust proceeds down to $\sim10$~K so that thermal
pressure becomes negligible compared to other sources, such as
magnetic and turbulent pressure. A magnetized,
supersonically-turbulent, self-gravitating gas clump is prone to
fragmentation: stochastic shock compressions produce a range of
unstable core masses down to about the thermal Jeans mass. This
process is likely to be responsible for shaping the present-day
stellar IMF. Although Pop III.1 cores are permeated by mild (sonic)
turbulence, the compressions produced by these fluctuations are quite
weak, and, coupled with the weak cooling properties of primordial
composition gas, means that fragmentation is suppressed.  As a result,
the IMF of Pop III.1 stars is likely to be qualitatively different
from that of Pop II and Pop I stars: it should be set by the
distribution of the entropy parameter $K'$ and the rotation parameter
$\fkep$.  Most numerical simulations of Pop III.1 star formation have
not exhibited fragmentation in the minihalo down to their resolution
limits, unless it occurs in a rotationally supported disk. Even this
latter case appears to be unlikely to occur for realistic cosmological
initial conditions.

A second difference is that the sites of Pop III.1 star formation are
determined by dark matter potentials, i.e. the centers of
minihalos. The cooling and settling of baryons mean that these
typically dominate the mass densities on scales inside $\sim 1$ ~pc,
but the dark matter still affects the gravitational potential of the
larger scale core, which can affect the efficiency of feedback
mechanisms. Dark matter annhilation in the centers of the halos could
provide an important heating mechanism that could delay the onset of
star formation \cite{spolyar2007}.

\subsection{Feedback Mechanisms}

One may ask how the feedback mechanisms of Pop III.1 stars relate to
those that operate in contemporary massive star formation. We note
that the maximum mass attained in the fiducial model of McKee \& Tan
of Pop III.1 star formation \cite{mckee2008} is very similar to that
inferred observationally in local massive star clusters
(e.g. \cite{figer2005}). However, after decades of study, it remains
unclear whether the maximum mass of stars forming today is set by
feedback or instabilities in very massive stars \cite{larson1971}.  We
have argued that the maximum mass of primordial stars is set by
feedback. The primary differences in the feedback processes then and
now are:

(1) Dust.  In contemporary star-forming regions, dust destroys \lal\
photons, eliminating them as a significant pressure. On the other
hand, the dust couples the pressure of the UV continuum radiation to
the gas very effectively, and it remains to be determined whether this
limits the final mass of the star; e.g., Yorke \& Sonnhalter
\cite{yorke2002} find that it does, whereas Krumholz et
al. \cite{krumholz2005a} have not found evidence that it does.  Dust
also affects the evolution of \ion{H}{2} regions, absorbing a
significant fraction of the ionizing photons in dense \ion{H}{2}
regions, thereby reducing the impact of \ion{H}{2} region breakout and
disk photoevaporation feedback.

(2) Magnetic fields. In contemporary protostars, magnetic fields drive
powerful winds that drive away a significant fraction of
the core out of which the star is forming \cite{matzner2000}. The
cavities created by these winds allow radiation to escape from the
vicinity of the protostar, significantly reducing the radiation
pressure \cite{krumholz2005b}. As we have discussed, it is
uncertain whether primordial protostars have such powerful
magnetically-driven outflows \cite{tan2004b}. However, even if they
are present, we have argued that they would not be the dominant
mechanism compared to disk photoevaporation in setting the final
stellar mass.

(3) Stellar temperatures and luminosities. Primordial stars were
significantly hotter than contemporary stars, resulting in
significantly greater ionizing luminosities. In addition, the
accretion rates of primordial massive stars are much greater, at least
initially, than those of contemporary massive stars \cite{mckee2003}.

\begin{theacknowledgments}
We thank many colleagues for helpful discussions, including T. Abel,
F. Adams, V. Bromm, A. Ferrara, K. Freese, G. Meynet, A. Natarajan,
B. O'Shea, E. Scannapieco, D. Spolyar, D. Whalen, S. Woosley and
N. Yoshida.  The research of JCT is supported by NSF CAREER grant
AST-0645412.  The research of CFM is supported by NSF grants
AST-0606831 and PHY05-51164.
\end{theacknowledgments}


\begin{thebibliography}{9}


\bibitem{page2007}
L. Page, G. Hinshaw, E. Komatsu et al., \emph{ApJS}, \textbf{170}, 335--376 (2007).

\bibitem{morales2004}
M. F. Morales, and J. Hewitt, \emph{ApJ}, \textbf{615}, 7--18 (2004).

\bibitem{beers2005}
T. C. Beers, and N. Christlieb, \emph{ARA\&A}, \textbf{43}, 531--580 (2005).

\bibitem{schaye2003}
J. Schaye, A. Aguirre, T-S. Kim et al., \emph{ApJ}, \textbf{596}, 768--796 (2003).

\bibitem{norman2004}
M. L. Norman, B. W. O'Shea, and P. Paschos, \emph{ApJ}, \textbf{601}, L115--L118 (2004).

\bibitem{santos2002}
M. R. Santos, V. Bromm, and M. Kamionkowski, \emph{MNRAS}, \textbf{336}, 1082--1092 (2002).

\bibitem{fernandez2006}
E. R. Fernandez, and E. Komatsu, \emph{ApJ}, \textbf{646}, 703--718 (2006).

\bibitem{kashlinsky2004}
A. Kashlinsky, R. Arendt, J. P. Gardner, et al., \emph{ApJ}, \textbf{608}, 1--9 (2004).

\bibitem{thompson2007}
R. I. Thompson, D. Eisenstein, X. Fan, et al., \emph{Apj}, \textbf{657}, 669--680 (2007).

\bibitem{weinmann2005}
S. M. Weinmann, and S. J. Lilly, \emph{ApJ}, \textbf{624}, 526--531 (2005).

\bibitem{bromm2002}
V. Bromm, and A. Loeb, \emph{ApJ}, \textbf{575}, 111--116 (2002).

\bibitem{stark2007}
D. P. Stark, R. S. Ellis, J. Richard, et al., \emph{ApJ}, \textbf{663}, 10--28 (2007).

\bibitem{fan2003}
X. Fan, M. A. Strauss, Schneider, D. P. et al., \emph{AJ}, \textbf{125}, 1649--1659 (2003).

\bibitem{willott2003}
C. J. Willott, R. J. McLure, and M. J. Jarvis, \emph{ApJ}, \textbf{587}, L15--L18 (2003).

\bibitem{mckee2008}
C. F. McKee, and J. C. Tan, \apj, submitted, astro-ph/0711.1377, (2008).

\bibitem{giroux1996}
M. L. Giroux, \& P. R. Shapiro, \emph{ApJS}, \textbf{102}, 191--238 (1996).

\bibitem{shapiro2004}
P. R. Shapiro, I. T. Iliev, \& A. C. Raga, \emph{MNRAS}, \textbf{348}, 753--782 (2004).

\bibitem{omukai2005}
K. Omukai, T. Tsuribe, R. Schneider, and A. Ferrara, \emph{ApJ}, \textbf{626}, 627--643 (2005)

\bibitem{bromm2003}
V. Bromm, and A. Loeb, \emph{Nature}, \textbf{425}, 812--814 (2003).

\bibitem{uehara2000}
H. Uehara, and S-I. Inutsuka, \emph{ApJ}, \textbf{531}, L91--94 (2000).

\bibitem{greif2006}
T. H. Greif, and V. Bromm, \emph{MNRAS}, \textbf{373}, 128--138 (2006).

\bibitem{tegmark1997}
M. Tegmark, J. Silk, M. J. Rees, A. Blanchard, T. Abel, and F. Palla, \emph{ApJ}, \textbf{474}, 1--12 (1997).

\bibitem{abn2002}
T. Abel, G. L. Bryan, and M. L. Norman, \emph{Science}, \textbf{295}, 93--98 (2002).

\bibitem{bcl2002}
V. Bromm, P. S. Coppi, and R. B. Larson, \emph{ApJ}, \textbf{564}, 23--51 (2002).

\bibitem{bromm2004}
V. Bromm, and A. Loeb, \emph{New Ast.}, \textbf{5}, 353--364 (2004).

\bibitem{tan2004}
J. C. Tan, and C. F. McKee, \emph{ApJ}, \textbf{603}, 383--400 (2004).

\bibitem{omukai1998}
K. Omukai, and R. Nishi 1998, \emph{ApJ}, \textbf{508}, 141--150 (1998).

\bibitem{ripamonti2002}
E. Ripamonti, F. Haardt, A. Ferrara, and M. Colpi, \emph{MNRAS}, {\bf 334}, 401--418 (2002).

\bibitem{hunter1977}
C. Hunter, C. \emph{ApJ}, \textbf{218}, 834--845 (1977).

\bibitem{shu1977}
F. H. Shu, \apj, \textbf{214}, 488--497 (1977).

\bibitem{yoshida2006}
N. Yoshida, K. Omukai, L. Hernquist, and T. Abel, \apj, \textbf{652}, 6--25 (2006).

\bibitem{o'shea2007}
B. W. O'Shea, and M. L. Norman, \apj, \textbf{654}, 66--92 (2007).

\bibitem{spolyar2007}
D. Spolyar, K. Freese, P. Gondolo, astro-ph/0705.0521 (2007).

\bibitem{stahler1986}
S. W. Stahler, F. Palla, \& E. E. Salpeter, \apj, \textbf{302}, 590--605 (1986). 

\bibitem{omukai2003}
K. Omukai, and F. Palla, \apj, \textbf{589}, 677--687 (2003).

\bibitem{schaerer2002}
D. Schaerer, \emph{A\&A}, \textbf{382}, 28--42 (2002).

\bibitem{gammie2001}
C. F. Gammie, \apj, \textbf{553}, 174--183 (2001).

\bibitem{adams1989}
F. C. Adams, S. P. Ruden, \& F. H. Shu, \apj, \textbf{347}, 959--976 (1989).

\bibitem{shu1990}
F. H. Shu, S. Tremaine, F. C. Adams, and S. P. Ruden, \apj, \textbf{358}, 495--514 (1990).

\bibitem{tan2004b}
J. C. Tan, and E. G. Blackman, \apj, \textbf{603}, 401-413 (2004).

\bibitem{balbus1998}
S. A. Balbus, and J. F. Hawley, Rev. Mod. Phys., \textbf{70}, 1--53 (1998).

\bibitem{kulsrud1997}
R. M. Kulsrud, R. Cen, J. P. Ostriker, \& D. Ryu, \apj, \textbf{480}, 481--491 (1997).

\bibitem{blackman2002}
E. G. Blackman, and G. B. Field, Phys. Rev. Lett., \textbf{89}, 265007 (2002).

\bibitem{shakura1973}
N. I. Shakura, and R. A. Sunyaev, A\&A, \textbf{24}, 337--355 (1973).

\bibitem{frank1995}
J. Frank, A. King, \& D. Raine, Accretion Power in Astrophysics (Cambridge: Cambridge Univ. Press) (1995).

\bibitem{artemova1996}
I. V. Artemova, G. S. Bisnovatyi-Kogan, G. Bjoernsson, \& I. D. Novikov, \apj, \textbf{456}, 119--123 (1996).

\bibitem{shu1992}
F. H. Shu, The Physics of Astrophysics Vol II: Gas Dynamics (Mill Valley: University Science Books) (1992).

\bibitem{iglesias1996}
C. A. Iglesias, and F. J. Rogers, \apj, \textbf{464}, 943--953 (1996).


\bibitem{hollenbach1994}
D. Hollenbach, D. Johnstone, S. Lizano, F. Shu, \apj, \textbf{428}, 654--669 (1994).

\bibitem{matzner2000}
C. D. Matzner, and C. F. McKee, \apj, \textbf{545}, 364--378 (2000).

\bibitem{matzner1999}
C. D. Matzner, and C. F. McKee, \apj, \textbf{526}, L109--L112 (1999).

\bibitem{machida2006}
M. N. Machida, K. Omukai, T. Matsumoto, S-I. Inutsuka, \apj, \textbf{647}, L1--L4 (2006).

\bibitem{tan2004c}
J. C. Tan, and C. F. McKee, in IAU Symp. 221, Star Formation at High Angular Resolution, ed. M. Burton, R. Jayawardhana, \& T. Bourke (San Francisco: ASP), astro-ph/0309139 (2004).


\bibitem{whalen2008}
D. Whalen, B. W. O'Shea, J. Smidt, M. L. Norman, \apj, submitted, astro-ph/0708.1603 (2008).

\bibitem{glover2001}
S. C. O. Glover, and P. W. J. L. Brand, MNRAS, \textbf{321}, 385--397 (2001).

\bibitem{susa2007}
H. Susa, \apj, \textbf{659}, 908--917 (2007).

\bibitem{ahn2007}
K. Ahn, and P. R. Shapiro, MNRAS, \textbf{375}, 881--908 (2007).

\bibitem{kudritzki2002}
R. Kudritzki, \apj, \textbf{577}, 389--408 (2002).

\bibitem{meynet2006}
G. Meynet, S. Ekstr\"om, A. Maeder, A\&A \textbf{447}, 623--639 (2006).

\bibitem{heger2002}
A. Heger, and S. E. Woosley, \apj, \textbf{567}, 532--543 (2002).

\bibitem{tumlinson2004}
J. Tumlinson, A. Venkatesan, and J. M. Shull, \apj, \textbf{612}, 602--614 (2004).

\bibitem{scannapieco2006}
E. Scannapieco, D. Kawata, C. B. Brook, R. Schneider, A. Ferrara,
B. K. Gibson, \apj, \textbf{653}, 285--299 (2006).

\bibitem{stacy2007}
A. Stacy, and V. Bromm, MNRAS, in press, astro-ph/0705.3634, (2007).

\bibitem{jappsen2007}
A. K. Jappsen, R.~S. Klessen, S. C. O. Glover, \& M.-M. Mac Low, astro-ph/0709.3530 (2007).

\bibitem{mueller2002}
K. E. Mueller, Y. L. Shirley, N. J. Evans, \& H. R. Jacobson, ApJS, \textbf{143}, 469--497 (2002).

\bibitem{figer2005}
D. F. Figer, Nature, \textbf{434}, 192--194 (2005).

\bibitem{larson1971}
R. B. Larson, and S. Starrfield, A\&A, \textbf{13}, 190--197 (1971).

\bibitem{yorke2002}
H. W. Yorke, and C. Sonnhalter, \apj, \textbf{569}, 846--862 (2002).

\bibitem{krumholz2005a}
M. R. Krumholz, C. F. McKee, R. I. Klein, in {\it Massive Star Birth:
A Crossroads of Astrophysics}, ed. R. Cesaroni, M. Felli,
E. Churchwell, \& M. Walmsley (Cambridge: Cambridge University Press),
p. 231 (2005).

\bibitem{krumholz2005b}
M. R. Krumholz, C. F. McKee, R. I. Klein, \apj, \textbf{618}, L33-36 (2005).

\bibitem{mckee2003}
C. F. McKee, and J. C. Tan, \apj, \textbf{585}, 850--871 (2003).


\end{thebibliography}
\end{document}